\documentclass{aa}  

\usepackage{graphicx}
\usepackage{txfonts}
\usepackage{amsmath}
\usepackage{threeparttable}
\usepackage{multirow}
\usepackage{booktabs} 
\usepackage{array}
\newcolumntype{R}[1]{>{\raggedleft\arraybackslash}m{#1}}
\usepackage[colorlinks=true, citecolor=blue, linkcolor = blue]{hyperref}

\begin{document} 

   \title{Clues to inside-out quenching in quiescent galaxies at $1.2\lesssim z\lesssim2.2$: Age, Fe-, and Mg-abundance gradients from \textit{JWST}-SUSPENSE}

   \author{Chloe M. Cheng
          \inst{1}\fnmsep\thanks{E-mail: cheng@strw.leidenuniv.nl (CMC)}
          \and
          Martje Slob\inst{1}
          \and
          Mariska Kriek\inst{1}
          \and
          Aliza G. Beverage\inst{2}
          \and
          Guillermo Barro\inst{3}
          \and
          Rachel Bezanson\inst{4}
          \and
          Anna de Graaff\inst{5}
          \and
          Natascha M. F\"orster Schreiber\inst{6}
          \and
          Brian Lorenz\inst{7}
          \and
          Danilo Marchesini\inst{8}
          \and
          Ignacio Martín-Navarro\inst{9,10}
          \and
          Adam Muzzin\inst{11}
          \and
          Andrew B. Newman\inst{2}
          \and
          Sedona H. Price\inst{12}
          \and
          Katherine A. Suess\inst{13}
          \and
          Arjen van der Wel\inst{14}
          \and
          Jesse van de Sande\inst{15}
          \and
          Pieter G. van Dokkum\inst{16}
          \and
          Daniel R. Weisz\inst{7}
          }

   \institute{Leiden Observatory, Leiden University, P.O. Box 9513, 2300 RA Leiden, The Netherlands
              \and
              Observatories of the Carnegie Institution for Science, 813 Santa Barbara Street, Pasadena, CA 91101, USA
              \and
              University of the Pacific, Stockton, CA 90340, USA
              \and
              Department of Physics \& Astronomy and PITT PACC, University of Pittsburgh, Pittsburgh, PA 15260, USA
              \and
              Max-Planck-Institut für Astronomie, Königstuhl 17, D-69117, Heidelberg, Germany
              \and
              Max-Planck-Institut f\"ur extraterrestrische Physik, Giessenbachstrasse 1, D-85748 Garching, Germany
              \and
              Department of Astronomy, University of California, Berkeley, CA 94720, USA
              \and
              Department of Physics \& Astronomy, Tufts University, MA 02155, USA
              \and
              Instituto de Astrofísica de Canarias,c/ Vía Láctea s/n, E38205 - La Laguna, Tenerife, Spain
              \and
              Departamento de Astrofísica, Universidad de La Laguna, E-38205 La Laguna, Tenerife, Spain
              \and
              Department of Physics and Astronomy, York University, 4700 Keele Street, Toronto, Ontario, ON MJ3 1P3, Canada
              \and
              Space Telescope Science Institute, 3700 San Martin Drive, Baltimore, MD 21218, USA
              \and
              Department for Astrophysical \& Planetary Science, University of Colorado, Boulder, CO 80309, USA
              \and
              Sterrenkundig Observatorium, Universiteit Ghent, Krĳgslaan 281 S9, B-9000 Gent, Belgium
              \and
              School of Physics, University of New South Wales, Sydney, NSW 2052, Australia
              \and
              Astronomy Department, Yale University, 52 Hillhouse Avenue, New Haven, CT 06511, USA
             }

   \date{Received XXX; accepted XXX}

\abstract
{Spatially resolved stellar populations of massive quiescent galaxies at cosmic noon provide powerful insights into star-formation quenching and stellar mass assembly mechanisms.  Previous photometric studies have revealed that the cores of these galaxies are redder than their outskirts. However, spectroscopy is needed to break the age-metallicity degeneracy and uncover the driver of these colour gradients. In this work, we derive the age and elemental abundance gradients for eight distant ($1.2 \lesssim z \lesssim 2.2$), massive ($10.3\lesssim\log({\rm M}_*/{\rm M}_\odot)\lesssim 11.1$) quiescent galaxies by fitting full-spectrum models to ultra-deep NIRSpec-MSA spectroscopy from the \textit{JWST}-SUSPENSE survey. We find that these galaxies have negative age and flat [Fe/H] gradients as well as tentative indications of positive [Mg/H] and [Mg/Fe] gradients. These results suggest that galaxy cores are older and perhaps also Mg deficient compared to galaxy outskirts. The age gradients may indicate inside-out quenching, while Mg-deficient cores could suggest rapid gas expulsion as the central quenching mechanism. Thus, galaxy cores may have formed faster and quenched more efficiently than their outskirts. In this scenario, however, our [Fe/H] and [Mg/Fe] gradients are still puzzling, and further investigation is required to understand the nature of [Mg/H] gradients in massive quiescent galaxies at these redshifts. Our results contrast with those of lower-redshift studies, which find flat age and [Mg/Fe] gradients and negative metallicity gradients.  Additionally, we find a positive trend between age gradients and rotational support and marginal trends between [Fe/H] gradients and galaxy velocity dispersions and ages.  We discuss our findings in the context of galaxy growth scenarios, including minor mergers and progenitor bias. With this work, we present the first stellar population gradients from NIRSpec-MSA spectroscopy in the current largest sample of distant quiescent galaxies.}


   \titlerunning{\textit{JWST}-SUSPENSE gradients}
   \maketitle

\section{Introduction}\label{sec:introduction}
Spatially resolved measurements of stellar populations are critical for our understanding of galaxy assembly histories and star-formation quenching mechanisms.  In the local Universe, key insights into the formation and assembly of massive early-type galaxies have been gained by measuring stellar population gradients out to large radii (e.g. \citealt{Peletier_1989, Franx_1990, Kuntschner_2006, Goddard_2017, Li_2018, Santucci_2020, Parikh_2024}).  In particular, findings of flat $\alpha$-element abundance gradients, flat or mildly positive age gradients, and mildly negative metallicity gradients (implying iron-rich cores) suggest that these nearby galaxies have experienced inside-out growth (e.g. \citealt{Mehlert_2003, La_Barbera_2005, Greene_2013, Greene_2015, Greene_2019, Martin_Navarro_2018, Zibetti_2020}). In this scenario, the compact central cores of massive galaxies are formed at cosmic noon ($z\sim 2-3$) or earlier, with their outer wings building up via minor mergers towards $z\sim0$ (e.g. \citealt{Bezanson_2009, Hopkins_2009, Naab_2009, Oser_2010, van_de_Sande_2013, van_dokkum_2014, Rodriguez_Gomez_2016}). 

Advancements in observations and modelling have allowed these measurements to be extended to higher redshifts ($z\sim1$).  In particular, negative colour gradients, (e.g. \citealt{Gargiulo_2012, Suess_2019a, Suess_2019b, Suess_2020, Suess_2021, Miller_2023, van_der_wel_2024}), flat age and $\alpha$-element abundance gradients, and negative metallicity gradients contribute to the inside-out growth picture (\citealt{Cheng_2024}; see also \citealt{D'Eugenio_2020}), with galaxy outskirts being built up by the accretion of low-metallicity (i.e. bluer) satellite galaxies.  In this context, however, the flat age and [Mg/Fe] gradients are more difficult to explain, as they require that the accreted satellites have similar ages and star-formation histories as the central galaxy.  On the other hand, we note that this may be expected for satellite galaxies (see e.g. \citealt{Pasquali_2010, Gallazzi_2021, Oyarzun_2023}).  

The resolved stellar populations of galaxies at even earlier times are crucial to assessing the inside-out growth picture.  High-resolution photometric studies beyond $z\sim1$ may in fact support this scenario, as colour gradients have been observed to strengthen between $z\sim2$ and $z\sim0$, with galaxies' blue outskirts building up over time (\citealt{Suess_2019b}, although note that this scenario is still under debate due to new evidence from JWST; see e.g. \citealt{Suess_2022, Martorano_2026, Mcgrath_2026}).  These colour gradients suggest that the compact cores of galaxies accrete bluer low-mass satellites (e.g. \citealt{Wuyts_2010, Guo_2011, Gargiulo_2012, Szomoru_2013, Ciocca_2017, Suess_2019b, Suess_2020, Suess_2021, Suess_2023, Miller_2022, Miller_2023, Setton_2024}).

On the other hand, the strengthening of colour gradients (as well as the increase in galaxy size) over cosmic time could be explained by larger galaxies with stronger gradients quenching at later times (i.e. progenitor bias; \citealt{van_dokkum_2001, Carollo_2013, Poggianti_2013, Keating_2015, Damjanov_2019, Damjanov_2023}).  In this case, gradients in galaxies at $z\gtrsim1$ would be inherited from their star-forming progenitors. This would suggest that both age and metallicity contribute to the colour gradients, as distant star-forming galaxies have been found to have negative metallicity (e.g. \citealt{Jones_2015}) and age gradients \citep{Tripodi_2024, Shen_2024}.  Colour gradients at $z\gtrsim1$ could also be signatures of inside-out quenching (e.g. \citealt{Suess_2019a}), as older central stellar populations result in redder cores.  Furthermore, dust could also play a role (e.g. \citealt{Miller_2023, Setton_2024}), as age, metallicity, and dust are strongly degenerate in broadband spectral energy distributions \citep{Worthey_1994, Bell_deJong_2001, Bruzual_2003, Gallazzi_2005, Leja_2019_sfh}. 

To be able to discriminate between the scenarios described above, the physical property that may be driving the observed colour gradients needs to be determined.  We must therefore measure robust stellar population gradients in massive quiescent galaxies out to at least $z\sim2$, the peak of the quenching and assembly era \citep{Oser_2010, Rodriguez_Gomez_2016, Whitaker_2012}.  To achieve this goal, deep, spatially resolved rest-frame optical spectra of a sample of massive quiescent galaxies are required to obtain detailed measurements of absorption lines.  However, these measurements are extremely challenging beyond the local Universe.  Thus, spectroscopic gradient measurements exist for only a small number of relatively young quiescent galaxies beyond $z\sim 1$.  For example, \cite{Jafariyazani_2020} found a flat age and [Mg/Fe] gradient and a slightly negative [Fe/H] gradient in a massive lensed galaxy at $z\sim2$ using deep MOSFIRE spectra.  \cite{Ditrani_2022} used \textit{Hubble Space Telescope} (\textit{HST}) grism spectra and found negative metallicity and diverse age gradients in four distant quiescent galaxies.  \cite{Akhshik_2023} examined similar data and found diverse age and metallicity gradients in eight galaxies.  However, due to line blending, metallicities obtained from low-resolution spectroscopy ($R\sim100$) may be highly uncertain ($\sim 0.3-0.5$ dex, \citealt{Akhshik_2023}).  

Detailed stellar population maps have also been achieved with \textit{James Webb Space Telescope} (\textit{JWST}) integral field unit (IFU) spectroscopy, but only for very young individual quiescent galaxies.  For example, \cite{Perez_Gonzalez_2024} examined a massive young (0.6 Myr old) galaxy at $z\sim3.7$ and found a strongly negative total metallicity gradient, a slightly negative age gradient, and a slightly positive dust gradient.  Additionally, \cite{D'Eugenio_2024} examined a young (0.5 Gyr old) galaxy at $z\sim 3$ and found a flat age gradient.  To come to a consensus about the behaviour of spatially resolved stellar populations in distant quiescent galaxies, larger, more representative samples that also include older, fainter galaxies are needed.  While \textit{JWST}-IFU observations such as those in \cite{Perez_Gonzalez_2024} and \cite{D'Eugenio_2024} are ideal, it is prohibitively expensive to obtain these data for larger samples. 

To avoid the expense of IFU observations, we can take advantage of the micro shutter assembly (MSA), \textit{JWST}'s multi-object spectrometer on the Near Infrared Spectrograph (NIRSpec).  This goal has been achieved with the \textit{JWST}-Spectroscopic Ultradeep Survey Probing Extragalactic Near-infrared Stellar Emission (SUSPENSE), an ultra-deep spectroscopic survey of 20 quiescent galaxies at $1 \lesssim z \lesssim 3$ \citep{Slob_2024}.  SUSPENSE leverages the NIRSpec-MSA to achieve both medium spectral and moderate spatial resolutions for many galaxies in a single pointing.  In this work, we took advantage of the unique capabilities of this instrument to present robust spatially resolved measurements of age, [Fe/H], [Mg/Fe], and [Mg/H] in a sample of eight massive quiescent galaxies at $1.2\lesssim z\lesssim 2.2$ from SUSPENSE. 

This paper is organised as follows: In Section~\ref{sec:data_sample} we describe the spectroscopic data from SUSPENSE and our sample selection.  In Section~\ref{sec:methods} we outline our methods to correct spectral resampling artefacts, extract spatially resolved spectra, determine de-projected distances, and measure spatially resolved stellar population parameters.  In Section~\ref{sec:results} we present our resulting gradients. We discuss our results and their implications for galaxy quenching mechanisms and assembly histories in Section~\ref{sec:discussion} and present our conclusions in Section~\ref{sec:conclusions}.  Throughout this work we assume a flat $\Lambda$ cold dark matter cosmology with $\Omega_m = 0.3$, $\Omega_\Lambda = 0.7$, and $H_0 = 70{\ }{\rm km}{\ }{\rm s}^{-1}{\ }{\rm Mpc}^{-1}$ and a \cite{Kroupa} initial mass function (IMF).  All magnitudes are given in the AB-magnitude system \citep{Oke_1983}.

\begin{table*}																							
	\centering																						
	\label{tab:sample}																						
	\begin{threeparttable}																						
	\caption{Sample of massive quiescent galaxies from \textit{JWST}-SUSPENSE examined in this work.}						\small																
	\begin{tabular}{m{2.2em}m{1.5em}m{4em}rrcm{3em}cm{2em}m{2em}m{2em}m{2em}} 																							
		\hline																							
        \multirow{3}{*}{ID} & \multirow{3}{*}{$z_{\rm spec}$\tnote{a}} &																							
        \multirow{3}{*}{$\log(M/M_\odot)$\tnote{a}} & \multicolumn{2}{c}{Kinematic Properties\tnote{b}} & \multicolumn{2}{c}{$R_{\rm e}$\tnote{c}} & \multirow{3}{*}{Axis ratio ($b/a$)\tnote{b}} & \multicolumn{2}{c}{Radius\tnote{c}} & \multicolumn{2}{c}{S/N$_{4520-4820\text{\AA}}$\tnote{d}} \\																							
        \cmidrule(lr){4-5}\cmidrule(lr){6-7}\cmidrule(lr){9-10}\cmidrule(lr){11-12}																							
        & & & $V_{r_{\rm e}}/\sigma_0$ & $\sigma_{0}$ & intrinsic\tnote{b} & convolved &  & core & outskirt & core & outskirt \\																							
        & & & & (km/s) &  (kpc) & (kpc) &  & ($R_{\rm e, conv}$) & ($R_{\rm e, conv}$) & (\AA$^{-1}$) & (\AA$^{-1}$)  \\																							
        \hline																							
130040	&	1.170	&	11.1	&	$0.63_{-0.07}^{+0.06}$	&	$285_{-5}^{+5}$	&	5.12\tnote{e}	&	5.57	&	0.53\tnote{e}	&	0.17	&	0.45	&	$21.2$	&	$19.9$	 \\
127345	&	1.171	&	10.7	&	-	&	-	&	$1.52\pm0.66$	&	2.13	&	0.78	&	0.40	&	0.95	&	$36.8$	&	$31.1$	 \\
127154	&	1.205	&	10.8	&	-	&	$242_{-6}^{+4}$	&	$2.12\pm0.41$	&	2.72	&	0.41	&	0.72	&	1.24	&	$54.2$	&	$30.7$	 \\
127108	&	1.335	&	10.3	&	$0.54_{-0.20}^{+0.12}$	&	$210_{-10}^{+5}$	&	$1.43\pm0.68$	&	2.02	&	0.75	&	0.51	&	0.97	&	$29.2$	&	$21.7$	 \\
129149	&	1.578	&	11.0	&	$0.49_{-0.12}^{+0.07}$	&	$387_{-34}^{+4}$	&	$1.00\pm0.08$	&	2.77	&	0.33	&	0.73	&	1.56	&	$91.7$	&	$52.7$	 \\
128041	&	1.760	&	10.7	&	$1.48_{-0.06}^{+0.07}$	&	$226_{-1}^{+1}$	&	$1.60\pm0.42$	&	2.77	&	0.45	&	0.75	&	1.01	&	$58.2$	&	$43.7$	 \\
129133	&	2.139	&	11.1	&	$1.19_{-0.07}^{+0.07}$	&	$257_{-1}^{+2}$	&	$1.58\pm0.83$	&	2.43	&	0.37	&	0.53	&	1.33	&	$59.1$	&	$29.1$	 \\
128036	&	2.196	&	11.0	&	$1.18_{-0.08}^{+0.09}$	&	$210_{-6}^{+3}$	&	$1.05\pm0.25$	&	1.85	&	0.65	&	0.55	&	1.09	&	$47.3$	&	$33.2$	 \\
\hline																																									
        \normalsize
	\end{tabular}																						
	\begin{tablenotes}																						
	\item[a] Presented in \cite{Slob_2024}.																						
	\item[b] Presented in \cite{Slob_2025}.																						
	\item[c] See Section~\ref{sec:radii}.	
    \item[d] See Section~\ref{sec:data_sample}.  Note that these S/N values are quoted per rest-frame \AA.
    \item[e] Presented in \cite{Griffith_2012}.
	\end{tablenotes}																						
	\end{threeparttable}																						
																							
\end{table*}																							

\section{Data and sample}\label{sec:data_sample}
\begin{figure*}
    \centering
    \includegraphics[width=0.8\textwidth]{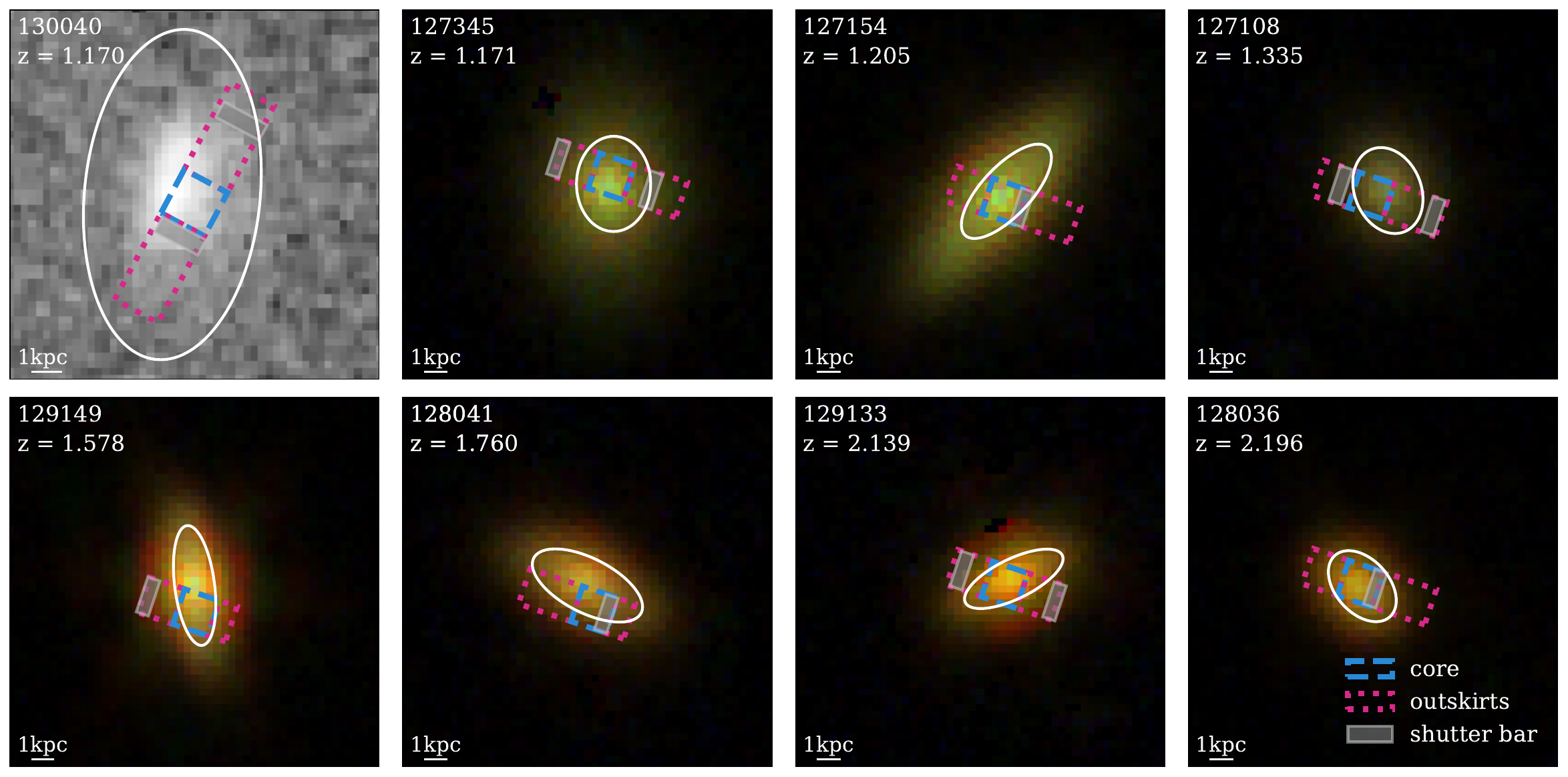}
    \caption{Colour image cut-outs of the eight distant quiescent galaxies in our sample.  Seven galaxies have NIRCam imaging, which we combine into RGB images here. Galaxy 130040 does not have NIRCam imaging, so we show a cut-out of the COSMOS \textit{HST}/ACS F814W image (\citealt{COSMOS_ACS_mosaics, Scoville_2007}; note that this image is rotated by 90 degrees compared to the NIRCam images).  We highlight the regions that we combined for the core and outskirt spectra in blue and pink, respectively (see Section~\ref{sec:extraction}).  We also indicate where shutter bars lie within these core and outskirt regions with shaded rectangles.  We indicate $1\ R_{\rm e}$ in convolved space for each galaxy with an ellipse (see Section~\ref{sec:radii}).}
    \label{fig:image_cutout_overview}
\end{figure*}
The spectroscopic data in this work were drawn from \textit{JWST}-SUSPENSE\footnote{\url{https://suspense.strw.leidenuniv.nl/}~.} (ID: 2110\footnote{\href{http://dx.doi.org/10.17909/y6rb-fn24}{doi:10.17909/y6rb-fn24}.}, PIs: Kriek and Beverage), an ultra-deep (16.4 hours on-source), NIRSpec-MSA/G140M-F100LP ($R\sim1300$) survey of 20 massive quiescent galaxies at $1 \lesssim z \lesssim 3$ (\citealt{Slob_2024}, see also \citealt{Beverage_suspense}).  Targets were identified from the UltraVISTA \citep{McCracken_2012} $K_s$-band DR3 catalogue \citep{Muzzin_2013}, and were selected to be quiescent using the \cite{Muzzin_2013_uvj} $UVJ$ classification.  The SUSPENSE galaxies were observed at two dither configurations, by offsetting two nearly identical MSA configurations by eight shutters in the dispersion direction, trying to maintain the same relative aperture pattern for each target.  However, due to micro shutter defects, this was not possible for all targets.  Thus, a small number of galaxies were observed in different positions in each configuration (see Section~\ref{sec:extraction}).  A two-point nod pattern was also implemented for each galaxy, with a two-micro-shutter cross-dispersion offset to avoid self-subtraction of the extended targets during the data reduction.  After identifying the optimal configuration, the standard three-shutter slitlets were extended by hand where possible, ranging from 3 to 7 micro shutters in length.  Filler star-forming targets were also added to the configuration.  The `unconstrained' (midbar) option was implemented for the source centring constraint, and thus some galaxies are centred behind MSA bars (see Fig.~\ref{fig:image_cutout_overview}). 

The data were reduced via a customised version of v1.12.5 of the \textit{JWST} Science Calibration Pipeline \citep{Bushouse_2023} and version 1183 of the Calibration Reference Data System (CRDS).  As we only make use of the reduced 2D spectra in this work (see Section~\ref{sec:extraction}), we summarise the relevant data reduction steps here, which were performed for each dither separately.  In particular, the master bias frame and dark current were subtracted and detector artefacts and jumps due to cosmic rays were removed.  Count-rate frames were obtained by fitting the slope of each pixel and $1/f$ correlated vertical read out noise was then removed using the \textsc{grizli} correction algorithm \citep{grizli}.  The data were background subtracted using the average of all frames observed in the same visit and dither but in the opposite nod.  Flat field, barshadow, and pathloss corrections were applied to the 2D spectra for each galaxy, which were then flux-calibrated.  The calibrated 2D spectra were rectified and resampled to a common reference frame.  A custom outlier detection algorithm was implemented, and the final 2D spectrum was constructed by performing an inverse-read-noise-weighted combination of the 2D frames.  See \cite{Slob_2024} for details. 

We selected a subset of the quiescent galaxies from SUSPENSE.  In particular, we required each spectrum (integrated, core, and outskirts, see Section~\ref{sec:extraction}) to have a rest-frame S/N $\gtrsim 20$ \AA$^{-1}$ between $4520-4820$ \AA\ (the region that all spectra have in common) to derive robust stellar population parameters.  We determined these limits by performing tests with simulated observations, similar to \cite{Cheng_2024} (see Appendix~\ref{sec:mocks}).  Additionally, we discarded four galaxies, due to insufficient wavelength coverage (IDs 129982 and 130208), or due to the presence of strong emission lines that may be associated with active galactic nuclei (AGN, IDs 128452 and 130647, see also \citealt{Beverage_suspense}).  Our selection does not lead to a bias in mass, size, or integrated age compared to the total SUSPENSE quiescent sample.    

Image cut-outs of our final sample of eight quiescent galaxies (40\% of the total quiescent SUSPENSE sample) are shown in Fig.~\ref{fig:image_cutout_overview}.  Seven galaxies have \textit{JWST}/Near Infrared Camera (NIRCam) imaging from the Cosmological Evolution Survey (COSMOS)-Web survey \citep{cosmos-web}.  We obtained these images from the Cosmic Dawn Centre (DAWN) \textit{JWST} Archive (DJA\footnote{\url{https://dawn-cph.github.io/dja/index.html}~.}, see \citealt{grizli} and \citealt{Valentino_2023}) and combined the F115W, F277W, and F444W data into RGB images using the \textsc{astropy} \texttt{make\_lupton\_rgb} function (this applies an inverse hyperbolic sine scaling, \citealt{Lupton_2004, astropy:2022}).  Galaxy 130040 does not have NIRCam imaging, so we show the COSMOS \textit{HST}/Advanced Camera for Surveys (ACS) F814W image \citep{COSMOS_ACS_mosaics, Scoville_2007}.  We report properties of our sample in Table~\ref{tab:sample}.

\section{Methods}\label{sec:methods}
\begin{figure*}
    \centering
    \includegraphics[width=\textwidth]{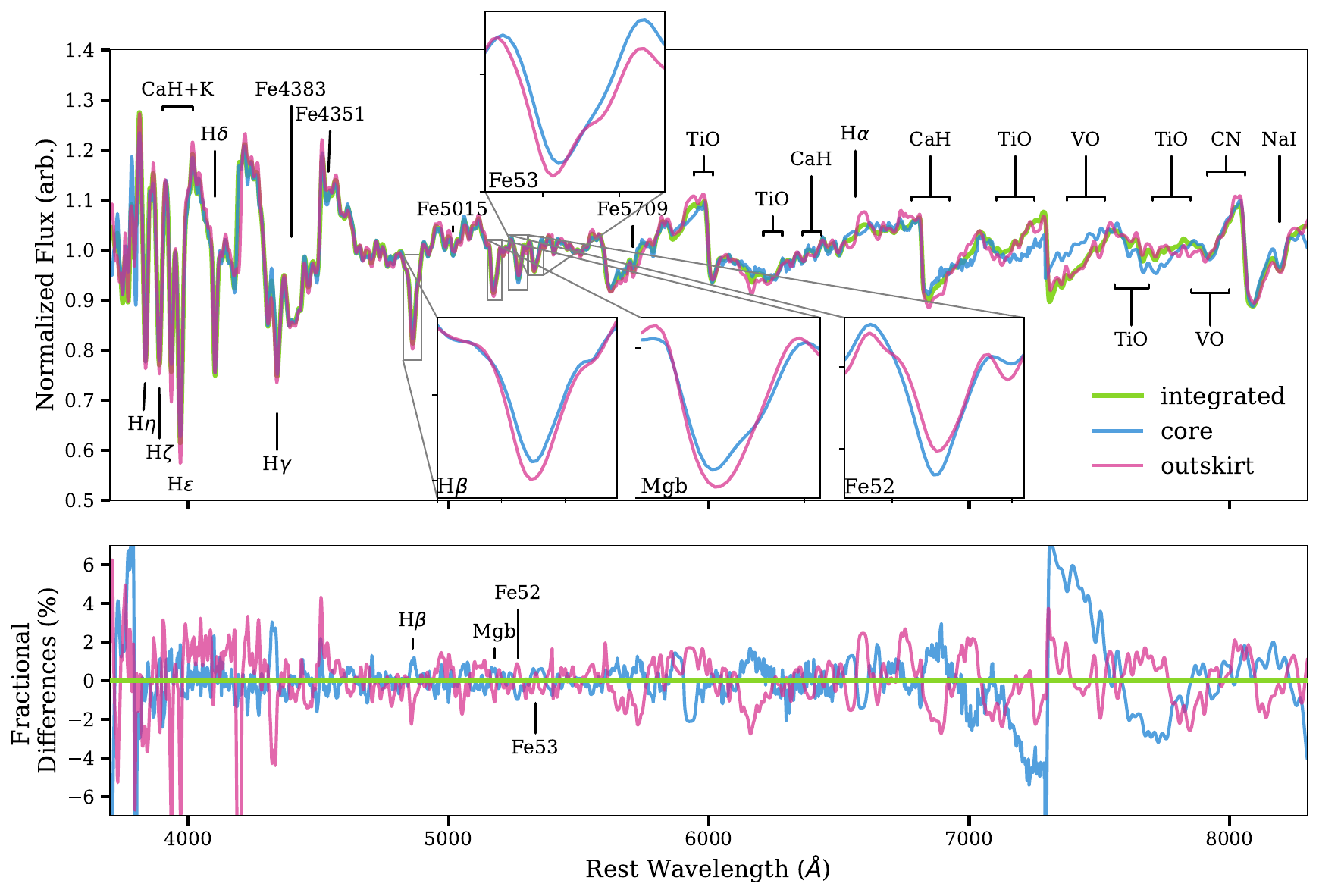}
    \caption{\textit{Top}: Stack of continuum-normalised spectra of all galaxies. The integrated stack is shown in green, the core stack is shown in blue, and the outskirt stack is shown in magenta.  The key spectral features are labelled.  The inset panels show a zoom-in on the H$\beta$ ($\sim 4849-4878$ \AA, sensitive to age), Mgb ($\sim 5162 - 5194$ \AA, sensitive to [Mg/H]), Fe52 ($\sim 5247 - 5287$ \AA, sensitive to age and [Fe/H]), and Fe53 ($\sim 5314 - 5454$ \AA, sensitive to age and [Fe/H]) features in the core and outskirt spectra.  \textit{Bottom}: Fractional differences of each stacked spectrum. We divided each spectrum by the stacked integrated spectrum.}
    \label{fig:all_stacked_spectra}
\end{figure*}

\subsection{Correction of resampling noise}\label{sec:wiggles}
Spatially resolved observations obtained with \textit{JWST} are affected by undersampling of the point spread function (PSF).  This undersampling produces an artefact called `resampling noise', which presents itself as periodic flux variations as a function of wavelength \citep{Smith_2007, Law_2023, Perna_2023, wicked, Newman_2025}.  These variations are colloquially referred to as `wiggles'.  This behaviour has been observed in the previous generation of space-based telescopes (see e.g. \citealt{Dressel_2007, Smith_2007, Anderson_2016}), and in \textit{JWST} IFU instruments (e.g. \citealt{Law_2023, Perna_2023, wicked, Newman_2025}).  We also observed these wiggles in the individual spectral rows of the SUSPENSE NIRSpec-MSA 2D spectra.  The wiggles do not strongly affect integrated observations, as they are averaged out when the flux is integrated over large apertures (see Appendix~\ref{sec:appendix_wiggles} and \citealt{Law_2023, Perna_2023, Newman_2025}).  However, the wiggles are a significant concern for spatially resolved studies as they can affect the shapes of continua and broad spectral features, potentially biasing spectral modelling \citep{Perna_2023}.  We corrected the wiggles by implementing an algorithm based on those presented by \cite{Perna_2023}\footnote{\url{https://github.com/micheleperna/JWST-NIRSpec_wiggles}~.} and \cite{wicked}\footnote{\url{https://github.com/antoinedumontneira/WiCKED}~.}, customising these routines for our specific case.  See Appendix~\ref{sec:appendix_wiggles} for details. 

\subsection{Spectral extraction}\label{sec:extraction}
For each galaxy, we applied a custom extraction routine to the wiggle-corrected 2D spectrum to obtain three 1D spectra: an integrated spectrum and spectra in two spatial bins (representing the core and outskirts).  We followed a similar procedure as \cite{Cheng_2024}, but modified for our data.  We describe our extraction routine here.  

We obtained the flux profile of each galaxy by collapsing the 2D spectrum over the wavelength axis.  We fit a \cite{Moffat_1969} profile to each flux profile, as in \cite{Cheng_2024}.  We used the Moffat profile to identify spectral rows with significant flux (rows that are approximately between the third and 97th percentiles of the Moffat profile).  We disregarded all rows outside of the third and 97th percentiles, as \cite{Cheng_2024} find that using this region results in a minimum difference in S/N between the core and outskirt spectra.  We also found that including these outer rows lowers the S/N. 

To extract the integrated spectrum, we summed the rows between the third and 97th percentiles.  For the spatially binned spectra, we took the brightest two rows to comprise the `core' of the galaxy and summed these together.  The remaining rows comprised the `outskirts' of the galaxy and were similarly summed.  We divided our spectra in this way to ensure that the two spatial bins had roughly similar and sufficient S/N.  Additionally, we chose this division to avoid splitting our rows into sub-pixels, since the spatial undersampling of NIRSpec makes neighbouring rows largely (though not completely) independent.  Table~\ref{tab:sample} shows that all core and outskirt bins have a median S/N $\gtrsim20$ \AA$^{-1}$, and that the S/N is roughly equally divided.  Finally, we did not apply an optimal extraction, so as to not down weigh the lower-signal outer rows.   

As the SUSPENSE galaxies were observed using two dither configurations (see Section~\ref{sec:data_sample} and \citealt{Slob_2024}), there are two 2D spectra for each galaxy.  We performed the extraction routine described above on the 2D spectrum in each dither, giving us an integrated, core, and outskirt spectrum for each dither.  We masked outliers in the flux due to cosmic ray impacts or bad pixels using the outlier detection algorithm from \cite{Slob_2024}.  For seven galaxies, the dithers encompass the same spatial regions.  In these cases, we mean-stacked the integrated, core, and outskirt spectra for each dither, respectively, leaving us with three 1D spectra (one integrated, core, and outskirt spectrum for each galaxy).  As a consequence of the MSA geometry, the two dithers for the remaining galaxy (130040) were performed in different positions on the galaxy.  This is because its centre was exactly between two shutters, and thus each dither encloses a different spatial region.  Therefore, for 130040 only, we analysed the core and outskirt spectra for each dither separately.  We found that the S/N in the outskirts of dither 1 was too low to robustly fit stellar population parameters (see Section~\ref{sec:data_sample} and Appendix~\ref{sec:mocks}).  This lower S/N could be due to astrometric errors, implying that the slit in dither 2 was better centred on the galaxy than in dither 1 and than indicated by the image in Fig.~\ref{fig:image_cutout_overview}.  Thus, we only considered dither 2 for this galaxy.  We show our defined core and outskirts regions for each galaxy in Fig.~\ref{fig:image_cutout_overview} (with only the dither 2 regions shown for 130040).

We show the stacked integrated, core, and outskirt spectra in Fig.~\ref{fig:all_stacked_spectra}.  We stacked all of our extracted, continuum-normalised (where we divided by a seventh order polynomial), rest-frame integrated, core, and outskirt spectra, respectively, by resampling the spectra to the same wavelength array via \textsc{spectres} \citep{SPECTRES}, normalising them by the region between $5200-5600$ \AA, and smoothing them to the maximum velocity dispersion of our sample ($\sigma\sim390\ {\rm km}\ {\rm s}^{-1}$).  We also show the fractional differences in the bottom panel.  The three spectra have visible variations relative to each other\footnote{Interestingly, there are large differences between the spectra in some of the molecular bands between $\sim7000-8000$ \AA.  It is beyond the scope of this work to discuss these bands in detail.  This will be the subject of future work.}.  To emphasise these differences, we zoom in on the H$\beta$, Mgb, Fe52, and Fe53 strong absorption features.  The four features demonstrate significant differences between the core and outskirts.  The H$\beta$ feature is largely sensitive to age, indicating that the average core stellar population has a different age compared to the average outskirt stellar population.  The Fe52 and Fe53 features are mainly sensitive to [Fe/H] but are also affected by age due to the age-metallicity degeneracy \citep{Worthey_1994, Bruzual_2003, Gallazzi_2005}.  Interestingly, the differences between the core and the outskirt spectra are opposite in these two features (i.e. Fe52 is stronger in the core while Fe53 is stronger in the outskirts).  Meanwhile, the Mgb feature is primarily sensitive to [Mg/H], indicating that the Mg abundances of the core and outskirt stellar populations may be different.  We discuss this further in Section~\ref{sec:results}.  

\subsection{Determining distances and radii}\label{sec:radii}
To present our spatially resolved measurements as a function of radius, we accounted for several factors.  These included: (i) the convolved nature of the observations, (ii) the fact that our galaxies may be elliptical in shape, and (iii) the arbitrary location of the micro shutters on each galaxy.

First, our observations are blurred due to convolution with the NIRSpec PSF.  To account for this effect and ensure consistency with the parameters derived from the PSF-convolved observations, we derived the convolved $R_{\rm e}$ ($R_{\rm e, conv}$) for each galaxy.  We used a similar method to those presented in \cite{Price_2016} and \cite{Cheng_2024}, where the intrinsic $R_{\rm e}$ ($R_{\rm e, int}$), measured by fitting images with \textsc{galfit} \citep{galfit} is also convolved with the NIRSpec PSF.  Specifically, we took the structural \textsc{galfit} parameters measured from \textit{JWST}/NIRCam F150W images by \cite{Slob_2025} (or those measured from COSMOS \textit{HST}/ACS F814W images by \citealt{Griffith_2012} for 130040) and used these parameters to create an intrinsic galaxy image in \textsc{galfit} for each galaxy.  We created these images at an increased spatial resolution (0.05\arcsec per pixel) compared to our observations.  We convolved each mock galaxy image with the NIRSpec-MSA PSF, generated with \textsc{stpsf}\footnote{\url{https://stpsf.readthedocs.io/en/latest/}~.} \citep{stpsf}, keeping all shutters open\footnote{Ideally, we would generate a PSF with the exact number of shutters open as in the SUSPENSE observations.  However, \textsc{stpsf} currently only offers simulated PSFs with one shutter, three adjacent shutters, or all shutters open.  We tested generating PSFs with three shutters open.  Our resulting convolved $R_{\rm e}$ measurements were consistent with keeping all shutters open, and thus a simulated PSF with the same number of shutters open as the observations is not likely to impact our conclusions.}.  We performed aperture sums on a series of elliptical apertures with increasing major and minor axes that we placed on the convolved model image using \textsc{photutils} \citep{photutils}, maintaining the intrinsic axis ratio measured with \textsc{galfit} in \cite{Slob_2025} (or \citealt{Griffith_2012} for 130040).  We determined the major axis enclosing 50\% of the light for the convolved model and took this to be our $R_{\rm e, conv}$.  We report $R_{\rm e, conv}$ for each galaxy in Table~\ref{tab:sample}.  

\begin{figure*}
    \centering
    \includegraphics[width=\textwidth]{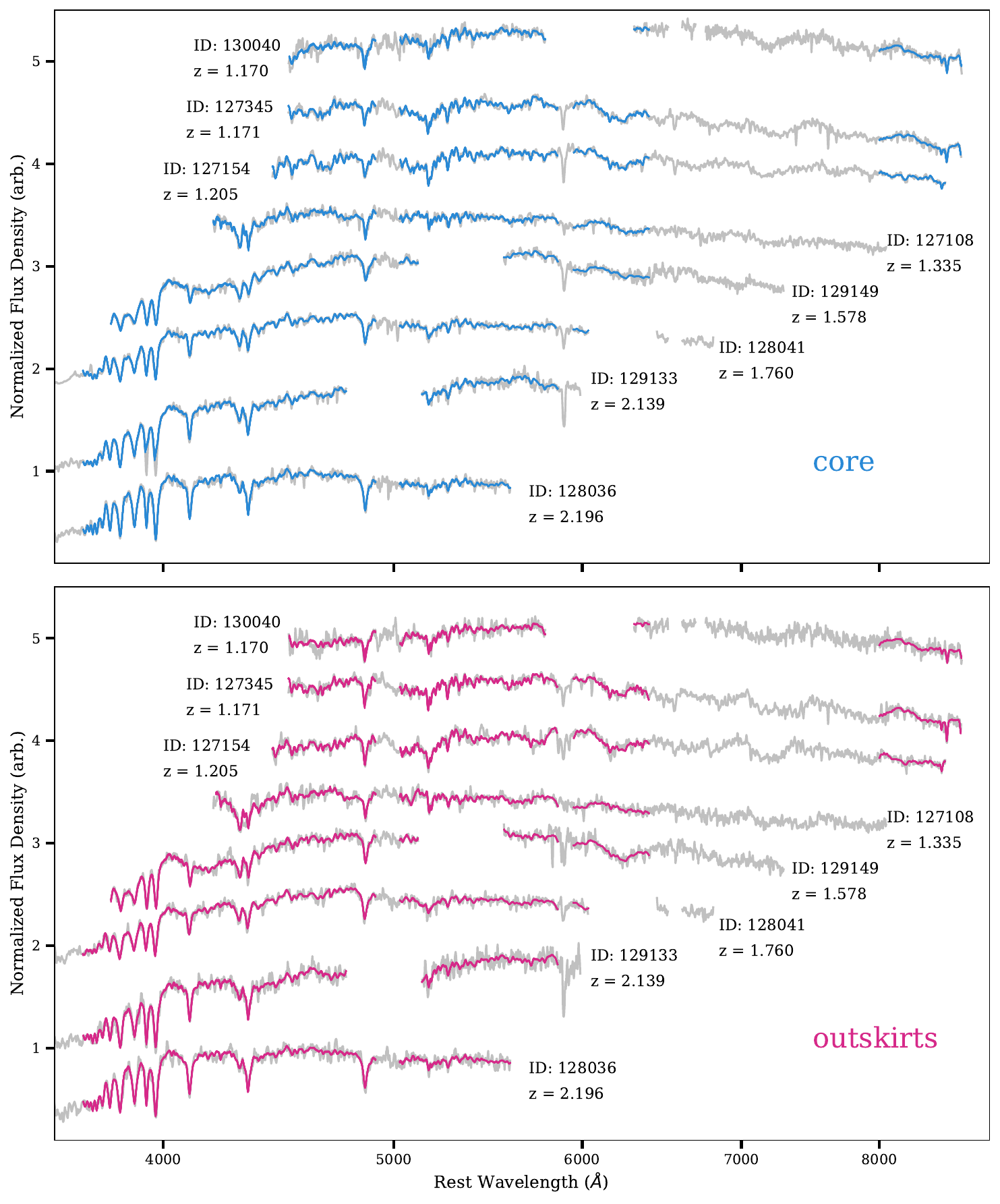}
    \caption{Best-fitting \textsc{alf$\alpha$} models to our quiescent galaxy spectra.  The core spectra (grey lines) are shown in the top panel in order of increasing redshift from top to bottom, with their best-fit models overplotted (blue lines).  The outskirt spectra (grey lines) and fits (magenta lines) are similarly shown in the bottom panel. We normalised each spectrum by its median value between $4480-4520$ \AA\ and arbitrarily offset them in the $y$-direction for visibility.  We do not show emission lines.  This figure illustrates that many absorption lines are robustly detected for both the core and outskirt regions.}
    \label{fig:fit_example}
\end{figure*}

\begin{table*}																											
	\centering																										
	\caption{Results from our \textsc{alf$\alpha$} fits to the spatially resolved \textit{JWST}-SUSPENSE spectra.}				\small								
	\label{tab:fit_results}																										
	\begin{tabular}{
        >{\raggedleft\arraybackslash}m{2.1em}
        >{\raggedleft\arraybackslash}m{3em}
        >{\raggedleft\arraybackslash}m{3em}
        >{\raggedleft\arraybackslash}m{3em}
        >{\raggedleft\arraybackslash}m{3em}
        >{\raggedleft\arraybackslash}m{3.5em}
        >{\raggedleft\arraybackslash}m{3.5em}
        >{\raggedleft\arraybackslash}m{3em}
        >{\raggedleft\arraybackslash}m{3em}
        >{\raggedleft\arraybackslash}m{3em}
        >{\raggedleft\arraybackslash}m{3em}
        >{\raggedleft\arraybackslash}m{3em}
        >{\raggedleft\arraybackslash}m{3em}
        } 																										
		\hline																									
		        ID & \multicolumn{3}{c}{$\log$(Age/Gyr)} & \multicolumn{3}{c}{[Fe/H]} & \multicolumn{3}{c}{[Mg/Fe]} & \multicolumn{3}{c}{[Mg/H]} \\																									
		        \cmidrule(lr){2-4} \cmidrule(lr){5-7} \cmidrule(lr){8-10} \cmidrule(lr){11-13}																									
		        & core & outskirt & gradient & core & outskirt & gradient & core & outskirt & gradient & core & outskirt & gradient \\																									
		        \hline																									
130040	&	$0.50_{-0.22}^{+0.19}$	&	$0.43_{-0.17}^{+0.22}$	&	$-0.18_{-0.53}^{+0.59}$	&	$-0.35_{-0.23}^{+0.19}$	&	$-0.41_{-0.22}^{+0.20}$	&	$-0.16_{-0.64}^{+0.63}$	&	$-0.02_{-0.17}^{+0.23}$	&	$0.33_{-0.22}^{+0.20}$	&	$0.92_{-0.72}^{+0.70}$	&	$-0.37_{-0.22}^{+0.23}$	&	$-0.09_{-0.27}^{+0.23}$	&	$0.25_{-0.57}^{+0.57}$	 \\
127345	&	$0.54_{-0.19}^{+0.15}$	&	$0.35_{-0.08}^{+0.10}$	&	$-0.55_{-0.38}^{+0.39}$	&	$-0.08_{-0.17}^{+0.14}$	&	$-0.07_{-0.22}^{+0.12}$	&	$\phantom{-}0.01_{-0.49}^{+0.46}$	&	$\phantom{-}0.21_{-0.13}^{+0.12}$	&	$0.33_{-0.19}^{+0.15}$	&	$0.35_{-0.54}^{+0.53}$	&	$\phantom{-}0.13_{-0.23}^{+0.16}$	&	$\phantom{-}0.25_{-0.32}^{+0.15}$	&	$0.03_{-0.42}^{+0.42}$	 \\
127154	&	$0.66_{-0.17}^{+0.12}$	&	$0.13_{-0.09}^{+0.11}$	&	$-2.46_{-0.46}^{+0.48}$	&	$-0.06_{-0.09}^{+0.09}$	&	$\phantom{-}0.14_{-0.27}^{+0.15}$	&	$\phantom{-}0.96_{-0.79}^{+0.76}$	&	-	&	-	&	-	&	-	&	-	&	-	 \\
127108	&	$0.58_{-0.13}^{+0.07}$	&	$0.52_{-0.14}^{+0.15}$	&	$-0.41_{-0.44}^{+0.45}$	&	$-0.55_{-0.15}^{+0.14}$	&	$-0.39_{-0.22}^{+0.18}$	&	$\phantom{-}0.54_{-0.95}^{+0.95}$	&	$\phantom{-}0.26_{-0.20}^{+0.20}$	&	$0.59_{-0.23}^{+0.17}$	&	$1.27_{-1.01}^{+1.02}$	&	$-0.29_{-0.20}^{+0.19}$	&	$\phantom{-}0.20_{-0.26}^{+0.16}$	&	$1.13_{-0.79}^{+0.73}$	 \\
129149	&	$0.24_{-0.05}^{+0.08}$	&	$0.21_{-0.08}^{+0.09}$	&	$-0.11_{-0.21}^{+0.21}$	&	$-0.12_{-0.17}^{+0.12}$	&	$-0.25_{-0.23}^{+0.17}$	&	$-0.38_{-0.60}^{+0.61}$	&	-	&	-	&	-	&	-	&	-	&	-	 \\
128041	&	$0.14_{-0.04}^{+0.07}$	&	$0.09_{-0.06}^{+0.10}$	&	$-0.41_{-0.44}^{+0.45}$	&	$-0.02_{-0.16}^{+0.10}$	&	$-0.08_{-0.23}^{+0.14}$	&	$-0.33_{-1.19}^{+1.17}$	&	$\phantom{-}0.10_{-0.14}^{+0.16}$	&	$0.13_{-0.21}^{+0.22}$	&	$0.25_{-1.76}^{+1.72}$	&	$\phantom{-}0.06_{-0.20}^{+0.15}$	&	$\phantom{-}0.03_{-0.29}^{+0.22}$	&	$0.05_{-1.47}^{+1.42}$	 \\
129133	&	$0.15_{-0.05}^{+0.06}$	&	$0.03_{-0.06}^{+0.09}$	&	$-0.31_{-0.15}^{+0.16}$	&	$-0.20_{-0.24}^{+0.11}$	&	$-0.16_{-0.29}^{+0.17}$	&	$\phantom{-}0.12_{-0.52}^{+0.49}$	&	-	&	-	&	-	&	-	&	-	&	-	 \\
128036	&	$0.09_{-0.04}^{+0.04}$	&	$0.00_{-0.04}^{+0.07}$	&	$-0.31_{-0.13}^{+0.14}$	&	$-0.27_{-0.22}^{+0.10}$	&	$-0.27_{-0.30}^{+0.17}$	&	$\phantom{-}0.04_{-0.62}^{+0.59}$	&	$\phantom{-}0.18_{-0.17}^{+0.17}$	&	$0.30_{-0.25}^{+0.21}$	&	$0.43_{-0.86}^{+0.85}$	&	$-0.09_{-0.24}^{+0.14}$	&	$\phantom{-}0.03_{-0.36}^{+0.20}$	&	$0.08_{-0.70}^{+0.64}$	 \\

		\hline	
        \normalsize
	\end{tabular}	
    \begin{tablenotes}																						
	\item \textbf{Notes.} For integrated stellar population parameters from SUSPENSE, see \cite{Beverage_suspense}.  For the `gradient' columns, we report $\Delta$Parameter/$\Delta\log_{10}(R_{\rm e, conv})$.
	\end{tablenotes}	
\end{table*}																																																									

To determine the average distance to the centre for our core and outskirt regions in units of $R_{\rm e, conv}$, we also considered that our galaxies are elliptical in shape on the sky, and are not perfectly round.  Additionally, we took into account the arbitrary placement of the micro shutters on each galaxy, which are typically not centred or aligned along the major axis.  To achieve this, we first determined the location of each spectral row on the sky.  We interpolated the coordinates of each row, by assuming each row was represented by a rectangle of width $0.1$\arcsec (the MSA pixel scale) and height 0.2\arcsec (the open width of a micro shutter, \citealt{Ferruit_2022}), calculating the distance between each corner of the rectangle, and dividing this distance into $\sim45000$ subpixels.  We found that this number of subpixels was approximately where our calculated distances stabilised, but we note that the precise number of subpixels does not affect our conclusions.  We generated a series of ellipses with increasing $R_{\rm e, conv}$ and $b_{\rm conv}$ centred on the target galaxy, maintaining the axis ratio.  Note that we disregarded the effect of the PSF on the ellipticity, and that smaller, inner ellipses will become rounder than outer ellipses.  We determined the $R_{\rm e, conv}$ of each subpixel from the dimensions of the respective intersecting ellipse.  The $R_{\rm e, conv}$ of that pixel was then the median $R_{\rm e, conv}$ of all of the subpixels in the row.  To determine the average $R_{\rm e, conv}$ of each spatial bin, we took the weighted mean of the $R_{\rm e, conv}$ of the rows in each region (core and outskirts), using the flux of each row as the weights.  Finally, we took the mean of the $R_{\rm e, conv}$ in each of the two observed dithers to get the final de-projected distance to the core and outskirts in units of $R_{\rm e, conv}$.  The dithers are at approximately the same position on each galaxy (other than for 130040, see above), but we took the mean of the $R_{\rm e, conv}$ in each dither to account for minor positional deviations.  Throughout the rest of the paper, we note that while we refer to `core' and `outskirt' regions somewhat arbitrarily for convenience, we present our results in units of the $R_{\rm e, conv}$ that we define here.  Thus, `core' regions can be taken to represent stellar populations at smaller distances (normalised to $R_{\rm e, conv}$), while `outskirt' regions can be taken to represent stellar populations at larger distances (normalised to $R_{\rm e, conv}$).  						

\subsection{Full spectrum fitting}\label{sec:alf}
We fit the spectra using \textsc{alf$\alpha$}\footnote{\url{https://github.com/alizabeverage/alfalpha}~.} \citep{alfa, Beverage_suspense}, a Python version of the \textsc{absorption line fitter} (\textsc{alf}\footnote{\url{https://github.com/cconroy20/alf/tree/master}~.}, \citealt{CvD_2012a, Conroy_2018}).  The \textsc{alf} and \textsc{alf$\alpha$} models are built on empirical simple stellar populations, constructed using the Mesa Isochrones and Stellar Tracks (MIST, \citealt{MIST}) and the Spectral Polynomial Interpolator\footnote{\url{https://github.com/AlexaVillaume/SPI_Utils}~.} (SPI, \citealt{Villaume_2017}).  \textsc{alf} and \textsc{alf$\alpha$} make use of stellar libraries, including the Medium Resolution INT Library of Empirical Spectra (MILES, \citealt{Sanchez_Blazquez_2006}), the Extended Infrared Telescope Facility stellar library (E-IRTF, \citealt{Villaume_2017}), the \cite{Mann_2015} sample of M-dwarf spectra, and the theoretical C3K stellar library (see \citealt{CvD_2012a}).  To fit the spectra, \textsc{alf$\alpha$} removes the continuum, by fitting a high-order Chebyshev polynomial to the ratio of the data to the model, and implements \textsc{dynesty} \citep{dynesty} or \textsc{emcee} \citep{emcee} to sample the posteriors of 20 stellar population parameters.  \textsc{alf} and \textsc{alf$\alpha$} can fit spectra with wavelengths ranging from $3700 - 24000$ \AA\ and stellar populations older than 1 Gyr.  For details, see \cite{CvD_2012a}, \cite{Conroy_2018}, and \cite{Beverage_suspense}.  We note that \textsc{alf$\alpha$} has been thoroughly tested and shown to produce the same results as \textsc{alf} \citep{Beverage_suspense}.

Prior to fitting, we smoothed the \textsc{alf$\alpha$} models to the instrumental resolution of the NIRSpec-MSA observations, where we made use of the instrumental resolution derived via \textsc{msafit} \citep{msafit} by \cite{Slob_2024}.  \cite{Slob_2024} find that the wavelength dependence of the instrumental resolution is typically a factor of 1.3 better than pre-launch estimates from JDox.  We assumed this corrected JDox curve when fitting.  See \cite{Slob_2024} and \cite{Beverage_suspense} for details.  We also masked the NaD absorption feature (which can be affected by the interstellar medium, \citealt{CvD_2012a}), [OIII] lines, and H$\alpha+$[NII] complex, where present, as in \cite{Beverage_suspense}.  Similar to \cite{Beverage_suspense}, we excluded regions between 6400 - 8000 \AA\ due to broad TiO absorption in the spectra.  Thus, where possible, we fit the wavelength regions 3700 - 4700 \AA, 4700 - 5100 \AA, 5100 - 5800 \AA, 5800 - 6400 \AA, and 8000 - 8600 \AA.  Finally, in general, NIRSpec-MSA noise has been found to be underestimated (see e.g. \citealt{Maseda_2023}).  Thus, we multiplied our noise spectra by the jitter term (an average factor of 2.1) fit in \cite{Beverage_suspense} for each SUSPENSE galaxy, prior to fitting. 

In our implementation of \textsc{alf$\alpha$}, we used \textsc{dynesty} to sample the posteriors of the velocity offset, velocity dispersion, single stellar population-equivalent age, isochrone metallicity, ten elemental abundances (Fe, C, N, Mg, Na, Si, Ca, Ti, and Cr), Balmer emission line flux, emission line velocity and broadening, a shift in the $T_{\rm eff}$ of the fiducial isochrones, and an instrumental jitter term to account for over or underestimation of the observed uncertainties.  We assumed a \cite{Kroupa} IMF.  We set the upper limit of the age prior to be the age of the Universe at the redshift of each galaxy, plus 2 Gyr to allow for uncertainties.  We assumed the default priors for all other parameters.  We examined the output ages and ensured that they are $\geq$ 1 Gyr, as \textsc{alf$\alpha$} is only suitable for stellar populations older than 1 Gyr.    

We first fit the integrated spectrum, allowing for variation in all stellar parameters listed above.  For the core and outskirt spectra, however, we could not constrain all of the elemental abundances at this S/N (we determined this by fitting mock observations similar to \citealt{Cheng_2024}, see Appendix~\ref{sec:mocks}).  Thus, as in \cite{Cheng_2024}, we fixed the values of all abundances other than Fe and Mg to the values from the integrated fits in our core and outskirt fits.  In this way, we only allowed age, Fe, and Mg to vary so that we could accurately constrain these parameters.  We visually inspected the results for each individual galaxy to ensure that the posterior distributions for age, Fe, and Mg were well-sampled.  If these posteriors ran up against the priors, we extended the priors for [Z/H], [Fe/H], and [Mg/H] as needed by up to 0.2 dex.  If the re-fitted age, [Z/H], or [Fe/H] posteriors still ran up against the priors, we discarded these galaxies from our sample.  For two galaxies (127154 and 129133), we found that the [Mg/H] posteriors ran up against the extended priors in at least one of the three fits for these objects.  Thus, for these objects only, we also fixed the Mg abundance in the core and outskirt fits to the value recovered in the integrated fits, and we do not report their Mg gradients.  We also do not report the Mg gradient for 129149 as there are no detectable Mg features in the spectra of this galaxy.  We show our core and outskirt \textsc{alf$\alpha$} fits in Fig.~\ref{fig:fit_example}. 

We compared our integrated fit results with those of \cite{Beverage_suspense} for the eight galaxies that we have in common (not shown).  Our results are consistent within $\sim0.2\sigma$, despite differences in the treatment of the data (i.e. the wiggle correction, see Section~\ref{sec:wiggles}) and spectral extraction (see \citealt{Slob_2024}).  

\begin{figure}
    \centering
    \includegraphics[width=\columnwidth]{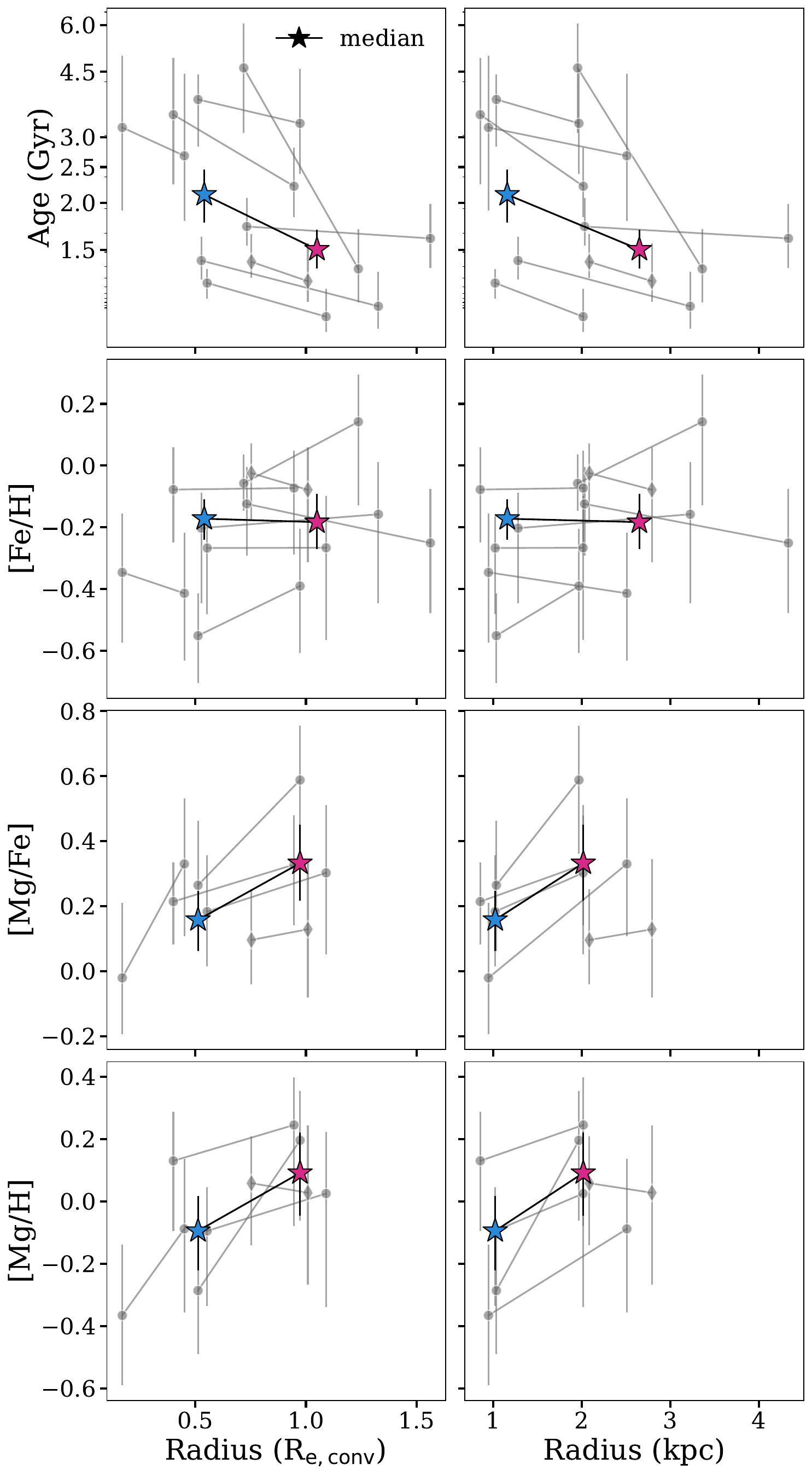}
    \caption{Spatially resolved stellar population parameters derived from our \textsc{alf$\alpha$} full spectrum fits.  In the left column, we plot our measured parameters as a function of de-projected, convolved radius in units of $R_{\rm e, conv}$ (determined as described in Section~\ref{sec:radii}).  In the right column, we plot our measured parameters as a function of de-projected, convolved radius in units of kpc.  We show age in the top row, [Fe/H] in the second row, [Mg/Fe] in the third row, and [Mg/H] in the bottom row.  In each panel, we show the measured parameters for each individual galaxy as points connected by lines.  For galaxy 128041, we note that the spectrum representing its core may partially probe its outskirt regions (see Fig.~\ref{fig:image_cutout_overview} and Table~\ref{tab:sample}) and show this galaxy as a diamond.  We show the median parameters of all galaxies as stars connected by lines.}
    \label{fig:params_vs_radius}
\end{figure}

\section{Results}\label{sec:results}			
\subsection{Spatially resolved ages and elemental abundances}\label{sec:individual_values}
In Fig.~\ref{fig:params_vs_radius}, we show the ages and elemental abundances in the cores and outskirts of eight massive quiescent galaxies at $1.2\lesssim z\lesssim 2.2$, derived by fitting the spectra with \textsc{alf$\alpha$}.  We show spatially resolved stellar population parameters as a function of de-projected radius from the centre of each galaxy.  We report these values in Table~\ref{tab:fit_results}.  We show galaxy 128041 as a diamond, as the spectrum representing its core may also partially probe its outskirt regions.  To estimate the uncertainties on the median values (stars), we performed a simple Monte Carlo simulation, where we took the median of the data points perturbed around the errors 1000 times, and we show the 16th and 84th percentiles of this distribution.    
 
\begin{figure*}
    \centering
    \includegraphics[width=0.7\textwidth]{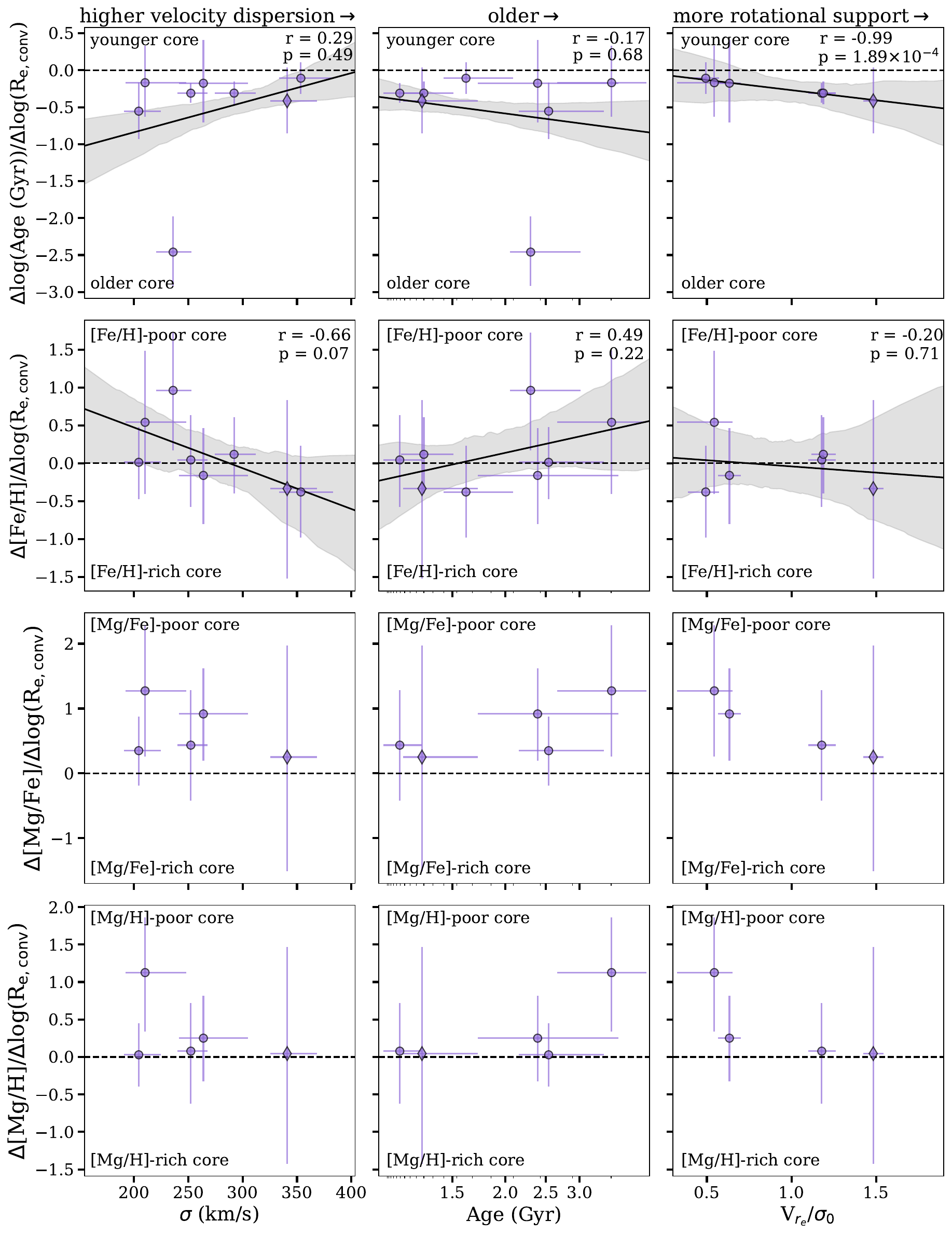}
    \caption{Spatially resolved stellar population gradients, normalised by $R_{\rm e, conv}$ (see Section~\ref{sec:radii}), as a function of galaxy parameters.  We show the age gradients in the top row, the [Fe/H] gradients in the second row, the [Mg/Fe] gradients in the third row, and the [Mg/H] gradients in the bottom row.  We indicate 128041 as a diamond (as in Figure~\ref{fig:params_vs_radius}), as its core spectrum partially extends into its outskirts.  The outlying object with a strongly negative age gradient is 127154.  We show the integrated velocity dispersion ($\sigma$, from our \textsc{alf$\alpha$} fits to the integrated spectra) in the left column, the integrated age (from our integrated \textsc{alf$\alpha$} fits) in the middle, and $V_{r_{\rm e}}/\sigma_0$ (measured in \citealt{Slob_2025}) on the right.  Horizontal dashed lines indicate where flat gradients would lie.  We performed a linear fit to the points in the age and [Fe/H] panels, shown by the solid lines, and $1\sigma$ bootstrapped uncertainties are indicated by the shaded regions.}
    \label{fig:grads_vs_params}
\end{figure*}

Fig.~\ref{fig:params_vs_radius} indicates that massive quiescent galaxies at $1.2 \lesssim z \lesssim 2.2$ from SUSPENSE tend to have negative age gradients, with galaxies having older cores compared to their outskirts.  Some galaxies may also have positive [Mg/H] gradients, with galaxy cores possibly being deficient in Mg compared to their outskirts (see Section~\ref{sec:trends} and Fig.~\ref{fig:grads_vs_params} as well).  We also find indications of positive [Mg/Fe] gradients, which may suggest that galaxy cores have longer SF timescales compared to galaxy outskirts (e.g. \citealt{Matteucci_1994, Maiolino_Mannucci_2019}, although note that the [Mg/H] and [Mg/Fe] gradients are overall consistent with being flat, within uncertainties).  We discuss this further in Section~\ref{sec:quenching}.  The individual SUSPENSE galaxies have diverse [Fe/H] gradients, although on average the [Fe/H] gradients are consistent with being flat, indicating that the cores and outskirts of these galaxies have similar metallicities.  The fact that our fits prefer the stellar populations at smaller $R_{\rm e, conv}$ to be older qualitatively agrees with the shallower H$\beta$ feature that we see in the inset panel in Fig.~\ref{fig:all_stacked_spectra}.  Similarly, the fact that we may observe stellar populations at smaller $R_{\rm e, conv}$ to have less Mg qualitatively agrees with the shallower Mgb feature.  Additionally, the contrasting differences in the core and outskirt spectra that we see in the Fe52 and Fe53 features could be contributing to our observed flat [Fe/H] gradient.  We discuss the implications of these results for galaxy formation and assembly in Section~\ref{sec:discussion}.

It is interesting to note that our results differ from gradients measured at lower $z$.  In particular, massive quiescent galaxies typically have flat age and [Mg/Fe] gradients, and negative metallicity gradients out to at least $z\sim1$ (e.g. \citealt{Mehlert_2003, Greene_2015, Martin_Navarro_2018, Cheng_2024, Parikh_2024}).  This suggests that stellar population gradients in massive quiescent galaxies may evolve from what we find at $1.2 \lesssim z \lesssim 2.2$ to what has been found at $z\lesssim1$.  In this context, it is interesting to note that the smallest quiescent galaxies in the local Universe (i.e. relic galaxies), have been found to have positive [Mg/Fe] gradients \citep{Martin_Navarro_2018}, potentially qualitatively similar to what we see here.  We discuss this further in Section~\ref{sec:discussion}.  

There are also a handful of other studies that have measured spatially resolved spectroscopic stellar population gradients in massive quiescent galaxies beyond $z\sim 1$.  Specifically, \cite{Jafariyazani_2020} measure a flat age and [Mg/Fe] gradient, and a negative [Fe/H] gradient in a massive lensed galaxy at $z = 1.98$, which is qualitatively consistent with results at $z\lesssim 1$.  \cite{Akhshik_2023} find diverse age gradients in eight lensed galaxies at $1.6 < z < 2.9$.  \cite{Ditrani_2022} measure age and metal gradients in four galaxies at $1.6 < z < 2.4$ and find flat age and negative metallicity gradients, again consistent with results at $z\lesssim 1$.  Finally, \cite{Perez_Gonzalez_2024} find a strong negative total metallicity gradient in an individual galaxy at $z\sim 3.7$ and \cite{D'Eugenio_2024} measure a constant age as a function of radius in an individual galaxy at $z\sim 3$.  It is not surprising that different works find diverse gradients beyond $z\sim 1$ as different methods are used in each of these studies.  Additionally, the sample sizes for which stellar population gradients can be measured at these redshifts (both in our work and in previous studies) are very small.  Finally, previous work examines relatively younger quiescent galaxies, which could explain some of these differences.  We discuss this further in Section~\ref{sec:trends}.

\subsection{Trends between gradients and galaxy parameters}\label{sec:trends}
As we have age and [Fe/H] gradients for our whole sample, we examined if there are any correlations with global velocity dispersions, integrated ages, and spatially resolved stellar kinematics.  These parameters are commonly used as measures of evolutionary stage.  In Fig.~\ref{fig:grads_vs_params}, we show the gradient slopes as a function of different galaxy parameters.  We fit a line to the points in the age and [Fe/H] panels using the non-linear least squares algorithm implemented in the \textsc{scipy} \texttt{curve\_fit} function, in the \texttt{optimize} module \citep{2020SciPy-NMeth}.  We computed the Pearson-$r$ correlation coefficient and corresponding $p$-value, which we state in each panel.  We note that we also show our [Mg/Fe] and [Mg/H] gradient slopes in the bottom two rows, but we did not fit a line to these points as we only have Mg abundances for 5 galaxies. 

In the left column, there is a negative correlation with [Fe/H] gradient.  Thus, galaxies with higher velocity dispersions have more negative [Fe/H] gradients.  We note, however, that further investigation is required to understand the significance of this apparent trend.  This tentative trend is in contrast with results at low-$z$ (see, e.g. \citealt{Sanchez_Blazquez_2007, Kuntschner_2010, Spolaor_2010, Pastorello_2014, Gonzalez_Delgado_2015, Greene_2015, Martin_Navarro_2018, Ferreras_2019, Santucci_2020}.  For example, \cite{Spolaor_2010} and \cite{Ferreras_2019} find positive correlations between metallicity gradients and mass/velocity dispersion, while \cite{Sanchez_Blazquez_2007}, \cite{Pastorello_2014}, \cite{Gonzalez_Delgado_2015}, and \cite{Greene_2015} find no dependence of metallicity gradients on mass.  

In the middle column of Fig.~\ref{fig:grads_vs_params}, there is a mild positive trend between [Fe/H] gradients and age, although this trend is also statistically insignificant.  Thus, younger quiescent galaxies may have more negative [Fe/H] gradients, while older quiescent galaxies may have more positive [Fe/H] gradients.  Interestingly, this tentative trend is qualitatively similar to the results of \cite{Cheng_2024}, who find a positive trend between [Fe/H] gradients and age in quiescent galaxies at $0.6\lesssim z\lesssim 1.0$.  However, we note that \cite{Cheng_2024} also find a negative trend between age gradients and age, which is inconsistent with our findings here (although we sample a slightly younger age range).

In this context, the tentative trends that we find between [Fe/H] gradients and age may be qualitatively consistent with previous spatially resolved studies at $z > 1$.  In particular, \cite{Jafariyazani_2020}, \cite{Perez_Gonzalez_2024}, and \cite{D'Eugenio_2020} all examine relatively young galaxies and find flat/slightly negative age gradients and negative metallicity gradients.  We may be seeing a similar effect in our youngest galaxies (middle column of Fig.~\ref{fig:grads_vs_params}). 

Finally, in the right column of Fig.~\ref{fig:grads_vs_params}, we compare our gradients to spatially resolved stellar kinematics measured in \cite{Slob_2025}.  In particular, \cite{Slob_2025} derive their kinematic measurements from the NIRSpec-MSA spectra using a forward modelling approach.  We show our stellar population gradients as a function of their $V_{r_{\rm e}}/\sigma_0$, where a higher value of $V_{r_{\rm e}}/\sigma_0$ indicates more rotational support.  We find a significant negative correlation between our age gradients and $V_{r_{\rm e}}/\sigma_0$.  Thus, galaxies with more rotational support may have more strongly negative age gradients (with their cores being much older than their outskirts).  On the other hand, \cite{Greene_2019} find no significant correlations between stellar population gradients and the ratio of rotational to dispersion support in massive local elliptical galaxies. We discuss our trends in Section~\ref{sec:discussion}.

\section{Discussion}\label{sec:discussion}  
\subsection{Implications for galaxy evolution}\label{sec:implications}
\begin{figure*}
    \centering
    \includegraphics[width=\textwidth]{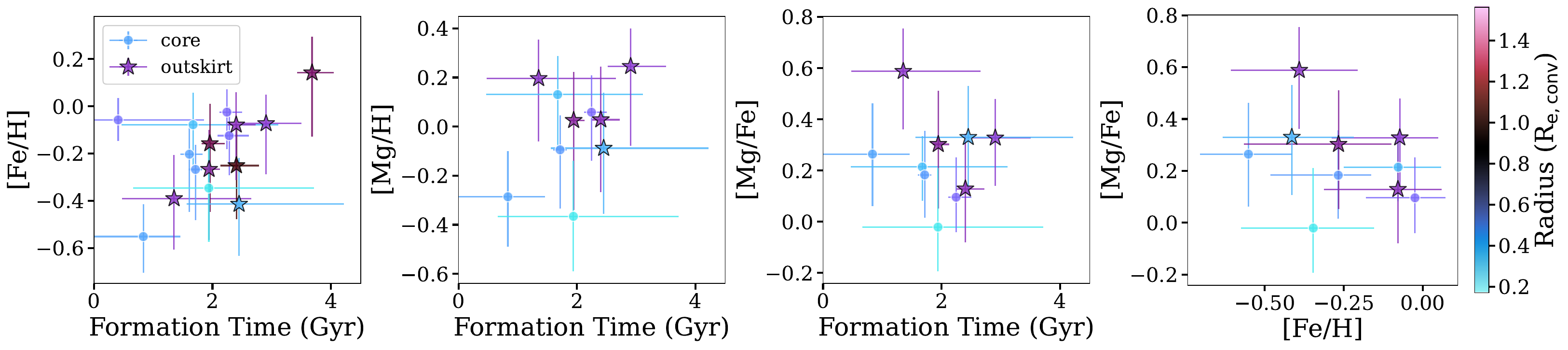}
    \caption{Individual core and outskirt elemental abundances.  In the first three panels, we show the formation time on the $x$-axis, which we computed by correcting the stellar age (from our \textsc{alf$\alpha$} fits to the integrated spectra) by the age of the Universe at the redshift of each galaxy.  We show our [Fe/H] abundances in the first panel, [Mg/H] abundances in the second panel, and [Mg/Fe] abundances in the third panel.  In the fourth panel, we show [Mg/Fe] as a function of [Fe/H].  In each panel, core values are shown as points, and outskirt values are shown as stars, and we colour-coded the symbols by their average distance to the centre of the galaxy in units of $R_{\rm e, conv}$.}
    \label{fig:formation_time}
\end{figure*}

Below $z\sim1$, resolved studies have revealed that massive quiescent galaxies tend to have flat age and $\alpha$-element abundance gradients and negative metallicity gradients (e.g. \citealt{Mehlert_2003, Greene_2015, Martin_Navarro_2018, Cheng_2024}).  These findings are consistent with inside-out growth, where outskirts are built up by the accretion of low-mass, low-metallicity satellites (minor mergers; e.g. \citealt{Bezanson_2009, Naab_2009, van_de_Sande_2013, Saputra_2025}).  Alternatively, the observed evolution in galaxy properties over cosmic time can be due to progenitor bias.  In this scenario, where the quiescent population grows by the addition of larger galaxies at later times \citep{van_dokkum_2001, Carollo_2013, Poggianti_2013}, gradients may come directly from the star-forming progenitors (\citealt{Cheng_2024}, see \citealt{Jones_2015, Tripodi_2024, Shen_2024, Ju_2025}), or could be produced by the quenching mechanism.  Here, we measured resolved stellar population gradients out to $z\sim2.2$.  Our galaxies were observed early, in the peak of the quenching and assembly era ($z\sim 1-3$; e.g. \citealt{Oser_2010, Rodriguez_Gomez_2016, Whitaker_2012}), and are more compact compared to galaxies at $z\lesssim 1$.  These characteristics suggest that they are relatively unpolluted by potential late-time mergers compared to nearby quiescent galaxies. Their gradients may thus provide interesting clues to quenching processes at $z\sim2$ and subsequent assembly processes towards $z\lesssim 1$.  

\subsubsection{Quenching at $z\sim2$}\label{sec:quenching}
The negative age gradients that we find may be the imprint of inside-out quenching, where star formation first occurs and/or terminates in the central regions of galaxies, followed by later-time star formation and/or quenching in the outskirts (e.g. \citealt{Dekel_Burkert_2014, Tacchella_2015a, Tacchella_2015b, Tacchella_2016b, Tacchella_2018, Zolotov_2015, Zibetti_2020}).  This scenario may be supported by our [Mg/H] gradients, which may be regulated by whether and how quickly the gas supply in the centres of the galaxies was depleted during the quenching process (see e.g. \citealt{Ellison_2018, Spilker_2019, Trussler_2020, Beverage_2021}).  In particular, \cite{Beverage_2021} compare different chemical evolution models from \cite{Spitoni_2017} and find that abrupt gas expulsion prevents the galaxy from further enriching, leading to lower metallicities for stellar populations that quench earlier, in agreement with observations (e.g. \citealt{Kriek_2016, Kriek_2019, Jafariyazani_2024, Beverage_2023, Beverage_2024, Beverage_suspense, Carnall_2022, Zhuang_2023}).  This scenario may thus explain our observed [Mg/H] abundances if the cores quenched more abruptly (perhaps due to AGN outflows; e.g. \citealt{Croton_2006, Bluck_2014, Zolotov_2015, Tacchella_2015b, Tacchella_2016b, Belli_2024, Park_2024}, although we note that our [Mg/H] gradients are also consistent with being flat within uncertainties).  

The inside-out quenching scenario presented here is reinforced by Fig.~\ref{fig:formation_time}, where we examine the general spatially resolved stellar populations in SUSPENSE as a function of formation time.  First, similar to \cite{Beverage_suspense}, we find that earlier-forming stellar populations may have slightly lower [Fe/H] and [Mg/H].  Moreover, stellar populations at smaller average distances from the centre of the galaxy tend to have formed earlier than stellar populations at larger average distances and may be less enriched by the time they stop forming stars, with tentative trends between [Fe/H] and distance from the centre.  Thus, this Figure reinforces the idea that the inner regions of the SUSPENSE galaxies may have formed earlier and also quenched more efficiently at earlier times.  

Although the scenario described above can explain the age gradients and perhaps some of the [Mg/H] gradients as well, it is puzzling that we find flat [Fe/H] gradients, as we would also expect more iron-deficient cores in this scenario.  Additionally, some of our [Mg/Fe] gradients may suggest that the star-formation timescale in the core is longer than in the outskirts (see e.g. \citealt{Matteucci_1994, Maiolino_Mannucci_2019}).  This is not what we expect from inside-out quenching, in the case that both populations started forming at the same time.  However, it is possible that star formation in the outskirts also started much later and occurred over a shorter time compared to the core.  It is also important to note that our measurements are luminosity-weighted, and thus our ages and [Mg/Fe] ratios, which may be driven by different wavelengths, may reflect slightly different stellar populations (although this effect may be minor).  Moreover, [Mg/Fe] is also influenced by the star-formation efficiency, the IMF, outflows, and late-time mergers (see e.g. \citealt{Tinsley_1979, Zolotov_2010, Martin_Navarro_2016, Andrews_2017, Sybilska_2018}).  In any case, our sample is small, and larger samples are required to come to definitive conclusions.

Finally, we note that our results are inconsistent with a scenario in which galaxies quench their star formation via wet, gas-rich, major mergers.  Major mergers are expected to result in outside-in quenching, by funnelling gas to the galaxy's centre and triggering a central starburst.  This would result in a positive age gradient \citep{Hopkins_2008, Snyder_2011, Wellons_2015, Pathak_2021}, which is generally inconsistent with our findings.  Nonetheless, it is possible that different quenching mechanisms operate at different redshifts.  For example, \cite{Cheng_2024} find positive age gradients in their youngest massive quiescent galaxies at $0.6\lesssim z\lesssim 1.0$, consistent with central starbursts triggered by wet mergers. 

\subsubsection{Galaxy evolution towards $z\lesssim 1$}\label{sec:assembly}
Assuming that the SUSPENSE galaxies are representative of the progenitors of $z\lesssim 1$ massive quiescent galaxies, it is interesting to compare our results with those from lower redshift studies (e.g. \citealt{Mehlert_2003, Greene_2015, Martin_Navarro_2018, Cheng_2024}).  In contrast to the flat age gradients at $z\lesssim 1$, we find negative age gradients at $1.2 \lesssim z\lesssim 2.2$.  In addition, while negative [Fe/H] gradients are found at lower redshifts, we find flat [Fe/H] gradients.  Finally, we find hints of positive [Mg/Fe] gradients, in contrast with the flat [Mg/Fe] gradients at $z\lesssim 1$.  Thus, age gradients may evolve from negative to flat, [Fe/H] gradients may evolve from flat to negative, and [Mg/Fe] gradients could evolve from positive to flat between $z\sim 2$ and $z\sim 1$.  This evolution could be due to inside-out growth via minor mergers, if the accreted galaxies are slightly older or have comparable ages to the central regions of distant quiescent galaxies, have lower [Fe/H], and possibly have lower [Mg/Fe].  In this scenario, the stars of these satellite galaxies will primarily be deposited in the outskirts and may flatten the age and [Mg/Fe] gradients that may have been generated by inside-out quenching, while simultaneously steepening the [Fe/H] gradients in the negative direction (see also \citealt{Zibetti_2020}).  We note that this scenario may not be fully consistent with the age-mass relation, where lower-mass galaxies are expected to be younger (e.g. \citealt{Kauffmann_2003}).  On the other hand, this expectation of lower-mass galaxies being younger may not hold for satellite galaxies (see, e.g. \citealt{Pasquali_2010, Gallazzi_2021, Oyarzun_2023}).   

This minor merger picture, which can also explain the difference in size between our galaxies and those at lower-$z$ (e.g. \citealt{Trujillo_2004, Franx_2008, Bezanson_2009, van_dokkum_2010, van_der_Wel_2014}) may be supported by some of the trends that we see with galaxy properties in Fig.~\ref{fig:grads_vs_params}.  Specifically, in the right column, we compare our resolved stellar population gradients to the six galaxies that we have in common with \cite{Slob_2025}, who measure spatially resolved stellar kinematics in 15 SUSPENSE galaxies.  Interestingly, we find a significant negative trend between our age gradients and $V_{r_{\rm e}}/\sigma_0$, where galaxies with more rotational support also have stronger negative age gradients.  This supports the scenario presented in \cite{Slob_2025}, where minor mergers gradually destroy rotational support over time while simultaneously flattening the age gradients (again, assuming that the accreted galaxies are older than the galaxy core).  However, larger sample sizes are required to constrain any trends between [Fe/H] gradients and $V_{r_{\rm e}}/\sigma_0$, and between age and [Fe/H] gradients and galaxy age, and accordingly assess whether these are also consistent with this picture (see also \citealt{D'Eugenio_2024}).   

Alternatively, progenitor bias \citep{van_dokkum_2001, Carollo_2013, Poggianti_2013} can also explain the evolution in gradients (as well as size) between the distant quiescent galaxies in our sample and those at $z < 1$.  In this scenario, the galaxies that quench at later times would have the stronger negative [Fe/H] and flat age gradients that we see at lower $z$.  These gradients could have been directly inherited from the star-forming progenitors and/or could indicate that different quenching processes operate at different times (see Section~\ref{sec:quenching}).    

For cases where galaxies are not affected by minor mergers, the high-$z$, quiescent galaxies must survive as they are until the present day.  Relic galaxies may represent this unique subpopulation, in which the early-forming cores of massive quiescent galaxies have passively evolved since $z\sim2$ without accreting \textit{ex situ} material \citep{Trujillo_2009, Quilis_2013, Stringer_2015}.  Thus, they may represent the descendants of distant quiescent galaxies, such as those in SUSPENSE.  Interestingly, and possibly similar to our findings, \cite{Martin_Navarro_2018} find positive [Mg/Fe] and flat metallicity gradients in a sample of nearby relic galaxies, which could have been directly inherited from the population of distant quiescent galaxies studied in this work.  On the other hand, our age gradients may be inconsistent with the finding of flat age gradients from \cite{Martin_Navarro_2018}.  Thus, it is likely that multiple mechanisms (i.e. progenitor bias and minor mergers) are playing a role (see also \citealt{van_der_wel_2008, Hopkins_2010, Oser_2012, Newman_2012}).

\subsection{Caveats}\label{sec:caveats}
In this work, we presented age and elemental abundance gradients in eight massive quiescent galaxies at $1.2\lesssim z \lesssim 2.2$.  We emphasise that this is the first study of its kind, with the largest sample size to date.  Thus, we have demonstrated that it is possible to measure stellar population gradients in galaxies beyond $z\sim1$ using the \textit{JWST}/NIRSpec-MSA.  However, there are some caveats that should be taken into consideration.  

There are a few observational factors affecting our data, including the limited spatial resolution and convolved nature of the MSA spectra, and the fact that the arbitrary position of the micro shutters does not always allow us to capture each galaxy's precise centre.  Despite our efforts to determine the exact regions of the galaxies contained within each micro shutter (see Section~\ref{sec:radii}), these factors likely result in some contamination from the outskirts in our core spectra and vice versa.  However, any contamination is likely minor, as we detect gradients in our galaxies.  This detection suggests that our measured gradients are real, as contamination from different regions of the galaxy will serve to weaken gradients (see \citealt{Cheng_2024}).  High spatial resolution follow-up with current or planned IFU instruments (i.e. the NIRSpec-IFU, ERIS on the VLT, HARMONI on the ELT) may provide a more detailed view of assembly histories by allowing measurements of detailed stellar population profiles (see e.g. \citealt{Oyarzun_2019}) and avoiding MSA slitlet offsets.  

Our fitting technique could also be biasing our results, as we only rely on one code (\textsc{alf$\alpha$}).  However, the stacked spectra that we show in Fig.~\ref{fig:all_stacked_spectra} are indicative of negative age and positive [Mg/H] gradients, with a shallower core H$\beta$ feature indicating an older age and a shallower Mgb feature potentially suggesting less [Mg/H] (although we note that Mgb is also sensitive to age).  Additionally, the fact that the Fe52 and Fe53 features show opposing core and outskirt depths may support a flat [Fe/H] gradient.  Thus, while the absolute values of our elemental abundances could be affected by our fitting technique, our qualitative gradients and our conclusions about them are presumably robust.  We note that we do not consider dust in our analysis.  However, in contrast to traditional stellar population synthesis (SPS) fitting results in the literature, our study does not suffer from degeneracies between dust and age/metallicity \citep{Worthey_1994, Bell_deJong_2001, Bruzual_2003, Gallazzi_2005}, as dust does not affect the depth of absorption features and our measurements are based on continuum-normalised spectra.

It is also possible that the IMF is contributing to the potential [Mg/Fe] gradient as elemental abundances and abundance ratios are dependent on the IMF (e.g. \citealt{Martin_Navarro_2016, Martin_Navarro_2018, De_Lucia_2017, Fontanot_2017}).  In this work, we assume a \cite{Kroupa} IMF, but there are indications that the low-mass end (e.g. \citealt{van_dokkum_2017, Gu_2022, Cheng_2026}) and perhaps also the high-mass end (e.g. \citealt{Fontanot_2017}, see also \citealt{van_dokkum_2024}) of the IMF in the SUSPENSE galaxies may deviate from the \cite{Kroupa} form (see also \citealt{Slob_2025}).  In particular, a variable IMF can affect the ratio of Mg to Fe by changing the relative number of core-collapse supernovae (which primarily produce Mg) compared to Type Ia supernovae (which primarily produce Fe, \citealt{Matteucci_1994, Thomas_1999}).  In general, it is challenging to constrain the shape of the IMF, as well as its impact on elemental abundances.  In the future, a robust characterisation of the slope of the IMF in distant quiescent galaxies, as well as detailed chemical evolution modelling to understand the effects of both star-formation timescales and the IMF on elemental abundance patterns, are required (see also \citealt{Beverage_suspense, Cheng_2026}). 

Finally, while this is the largest sample of $z\gtrsim 1$ galaxies for which gradients have been derived using this method so far, our sample size is still small, which makes it challenging to draw strong conclusions about the population of galaxies at these redshifts.  Thus, in the future, it will be informative to make the same measurements in a statistically significant sample of galaxies, perhaps with similar follow-up NIRSpec-MSA observations.   

\section{Summary and conclusions}\label{sec:conclusions}
Spatially resolving the stellar populations of massive quiescent galaxies at early times can reveal crucial information about how galaxies quenched their star formation and assembled their stellar populations.  Previous work beyond the local Universe has been based on photometry or very small spectroscopic sample sizes.  In this work, we leveraged the capabilities of the \textit{JWST}/NIRSpec-MSA and presented age, [Fe/H], [Mg/H], and [Mg/Fe] gradients in eight massive quiescent galaxies at $1.2 \lesssim z \lesssim 2.2$ from \textit{JWST}-SUSPENSE, an ultra-deep spectroscopic survey \citep{Slob_2024}.  These gradients were derived by fitting the central and outskirt spectra of each galaxy with \textsc{alf$\alpha$}, a flexible full-spectrum stellar population synthesis model \citep{CvD_2012a, Conroy_2018, Beverage_2024, Beverage_suspense}. Our study accounted for the fact that the NIRSpec-MSA micro shutters were not centred or aligned on the galaxies in our sample as well as for spatial undersampling of the \textit{JWST} PSF.  In summary, we found the following:
\begin{itemize}
    \item Massive quiescent galaxies at $1.2 \lesssim z \lesssim 2.2$ from SUSPENSE tend to have negative age gradients and flat [Fe/H] gradients, and they may have positive [Mg/H] and [Mg/Fe] gradients (although we note that the last two properties are also consistent with being flat within uncertainties). In other words, the cores of the SUSPENSE galaxies are generally older, possibly Mg enhanced, and have approximately the same [Fe/H] compared to their outskirts. 
    \item The negative age gradients suggest inside-out quenching (e.g. \citealt{Tacchella_2016b, Tacchella_2018}), where star formation terminated early in galaxy cores while galaxy outskirts continued to form stars until later times.  The possible low central [Mg/H] may indicate rapid central gas expulsion (e.g. \citealt{Beverage_2021, Beverage_suspense}), further supporting the inside-out quenching scenario.  Thus, galaxy cores formed earlier and may have stopped their star formation more abruptly, preventing further enrichment.  Meanwhile, galaxy outskirts formed later and may have experienced more enrichment before quenching.  However, it is unclear how our [Fe/H] and [Mg/Fe] gradients fit into this picture. 
    \item Our gradients differ from those at lower-$z$, where studies typically find flat age and [Mg/Fe] gradients and negative [Fe/H] gradients (e.g. \citealt{Greene_2015, Martin_Navarro_2018, Cheng_2024}).  Thus, gradients appear to evolve over cosmic time. 
    \item We examined the correlations between our gradients and galaxy properties including global velocity dispersion, integrated galaxy age, and $V_{r_{\rm e}}/\sigma_0$.  In particular, we found that age gradients are stronger in galaxies that have more rotational support.  This trend and the fact that gradients may evolve over time suggest that galaxy growth may continue post-quenching via minor mergers (e.g. \citealt{Bezanson_2009, Naab_2009, van_de_Sande_2013}).  In this scenario, the accretion of old low-mass and low-metallicity satellites could act to flatten age and [Mg/Fe] gradients and build up negative [Fe/H] gradients while simultaneously increasing galaxy sizes (e.g. \citealt{Trujillo_2004, van_dokkum_2010, van_der_Wel_2014}) and decreasing their rotational support \citep{Slob_2025}. 
    \item The observed `evolution' in gradients may also be explained by different quenching mechanisms occurring at different times, with early-time quenchers having negative age and flat [Fe/H] gradients and late-time quenchers having negative [Fe/H] and flat age gradients.  
\end{itemize}

In this work, we probed the peak of the star-formation quenching and galaxy assembly era with the first detailed stellar population gradients derived in a sample of distant massive quiescent galaxies using ultra-deep \textit{JWST}/NIRSpec-MSA spectra.  This study has allowed us to drive closer to the detailed formation histories of such galaxies, with our results suggesting that their cores formed earlier and quenched more abruptly than their outskirts.  However, there are still a number of open questions.  For example, it is not clear why we find indications of a positive [Mg/H] gradient but no gradient in [Fe/H].  It is also unclear how the IMF behaves in these galaxies and how it may affect the derived stellar population gradients.  In the future, larger sample sizes, increased spatial resolution, and a robust characterisation of the stellar IMF will paint an even clearer picture of the physical processes at play in the evolution of these galaxies.

\begin{acknowledgements}
We thank the anonymous referee for taking the time to give us useful feedback that improved this manuscript.  We thank Antoine Dumont, Ciarán Rogers, and Colin Yip for useful conversations.  We thank the \textit{JWST}/NIRSpec helpdesk for their support.  This work is based on observations made with the NASA/ESA/CSA \textit{JWST}.  The data were obtained from the Mikulski Archive for Space Telescopes (MAST) at the Space Telescope Science Institute (STScI), which is operated by the Association of Universities for Research in Astronomy, Inc., under NASA contract NAS 5-03127 for \textit{JWST}.  These observations are associated with program \textit{JWST}-GO-2110.  Support for program \textit{JWST}-GO-2110 was provided by NASA through a grant from the STScI.  Some of the data products presented herein were retrieved from the Dawn \textit{JWST} Archive (DJA). DJA is an initiative of the Cosmic Dawn Center (DAWN), which is funded by the Danish National Research Foundation under grant DNRF140.  This work was performed using the compute resources from the Academic Leiden Interdisciplinary Cluster Environment (ALICE) provided by Leiden University.  This work also used the Dutch national e-infrastructure with the support of the SURF Cooperative using grant no. EINF-10017 which is financed by the Dutch Research Council (NWO). MK acknowledges funding from the NWO through the award of the Vici grant VI.C.222.047.
\end{acknowledgements}

\bibliographystyle{aa}
\bibliography{references}

\begin{thebibliography}{167}
\expandafter\ifx\csname natexlab\endcsname\relax\def\natexlab#1{#1}\fi

\bibitem[{{Akhshik} {et~al.}(2023){Akhshik}, {Whitaker}, {Leja}, {Richard}, {Spilker}, {Song}, {Brammer}, {Bezanson}, {Ebeling}, {Gallazzi}, {Mahler}, {Mowla}, {Nelson}, {Pacifici}, {Sharon}, {Toft}, {Williams}, {Wright}, \& {Zabl}}]{Akhshik_2023}
{Akhshik}, M., {Whitaker}, K.~E., {Leja}, J., {et~al.} 2023, \apj, 943, 179

\bibitem[{{Anderson}(2016)}]{Anderson_2016}
{Anderson}, J. 2016, {Empirical Models for the WFC3/IR PSF}, Instrument Science Report WFC3 2016-12, 42 pages

\bibitem[{{Andrews} {et~al.}(2017){Andrews}, {Weinberg}, {Sch{\"o}nrich}, \& {Johnson}}]{Andrews_2017}
{Andrews}, B.~H., {Weinberg}, D.~H., {Sch{\"o}nrich}, R., \& {Johnson}, J.~A. 2017, \apj, 835, 224

\bibitem[{{Astropy Collaboration} {et~al.}(2022){Astropy Collaboration}, {Price-Whelan}, {Lim}, {Earl}, {Starkman}, {Bradley}, {Shupe}, {Patil}, {Corrales}, {Brasseur}, {N{"o}the}, {Donath}, {Tollerud}, {Morris}, {Ginsburg}, {Vaher}, {Weaver}, {Tocknell}, {Jamieson}, {van Kerkwijk}, {Robitaille}, {Merry}, {Bachetti}, {G{"u}nther}, {Aldcroft}, {Alvarado-Montes}, {Archibald}, {B{'o}di}, {Bapat}, {Barentsen}, {Baz{'a}n}, {Biswas}, {Boquien}, {Burke}, {Cara}, {Cara}, {Conroy}, {Conseil}, {Craig}, {Cross}, {Cruz}, {D'Eugenio}, {Dencheva}, {Devillepoix}, {Dietrich}, {Eigenbrot}, {Erben}, {Ferreira}, {Foreman-Mackey}, {Fox}, {Freij}, {Garg}, {Geda}, {Glattly}, {Gondhalekar}, {Gordon}, {Grant}, {Greenfield}, {Groener}, {Guest}, {Gurovich}, {Handberg}, {Hart}, {Hatfield-Dodds}, {Homeier}, {Hosseinzadeh}, {Jenness}, {Jones}, {Joseph}, {Kalmbach}, {Karamehmetoglu}, {Ka{l}uszy{'n}ski}, {Kelley}, {Kern}, {Kerzendorf}, {Koch}, {Kulumani}, {Lee}, {Ly}, {Ma}, {MacBride}, {Maljaars}, {Muna}, {Murphy}, {Norman}, {O'Steen},
  {Oman}, {Pacifici}, {Pascual}, {Pascual-Granado}, {Patil}, {Perren}, {Pickering}, {Rastogi}, {Roulston}, {Ryan}, {Rykoff}, {Sabater}, {Sakurikar}, {Salgado}, {Sanghi}, {Saunders}, {Savchenko}, {Schwardt}, {Seifert-Eckert}, {Shih}, {Jain}, {Shukla}, {Sick}, {Simpson}, {Singanamalla}, {Singer}, {Singhal}, {Sinha}, {Sip{H{o}}cz}, {Spitler}, {Stansby}, {Streicher}, {{{S}}umak}, {Swinbank}, {Taranu}, {Tewary}, {Tremblay}, {Val-Borro}, {Van Kooten}, {Vasovi{'c}}, {Verma}, {de Miranda Cardoso}, {Williams}, {Wilson}, {Winkel}, {Wood-Vasey}, {Xue}, {Yoachim}, {Zhang}, {Zonca}, \& {Astropy Project Contributors}}]{astropy:2022}
{Astropy Collaboration}, {Price-Whelan}, A.~M., {Lim}, P.~L., {et~al.} 2022, \apj, 935, 167

\bibitem[{{Bell} \& {de Jong}(2001)}]{Bell_deJong_2001}
{Bell}, E.~F. \& {de Jong}, R.~S. 2001, \apj, 550, 212

\bibitem[{{Belli} {et~al.}(2024){Belli}, {Park}, {Davies}, {Mendel}, {Johnson}, {Conroy}, {Benton}, {Bugiani}, {Emami}, {Leja}, {Li}, {Maheson}, {Mathews}, {Naidu}, {Nelson}, {Tacchella}, {Terrazas}, \& {Weinberger}}]{Belli_2024}
{Belli}, S., {Park}, M., {Davies}, R.~L., {et~al.} 2024, \nat, 630, 54

\bibitem[{{Beverage}(2024)}]{alfa}
{Beverage}, A. 2024, {alizabeverage/alfalpha: Initial Release alongside Beverage et al. 2024}

\bibitem[{{Beverage} {et~al.}(2021){Beverage}, {Kriek}, {Conroy}, {Bezanson}, {Franx}, \& {van der Wel}}]{Beverage_2021}
{Beverage}, A.~G., {Kriek}, M., {Conroy}, C., {et~al.} 2021, \apjl, 917, L1

\bibitem[{{Beverage} {et~al.}(2023){Beverage}, {Kriek}, {Conroy}, {Sandford}, {Bezanson}, {Franx}, {van der Wel}, \& {Weisz}}]{Beverage_2023}
{Beverage}, A.~G., {Kriek}, M., {Conroy}, C., {et~al.} 2023, \apj, 948, 140

\bibitem[{{Beverage} {et~al.}(2024){Beverage}, {Kriek}, {Suess}, {Conroy}, {Price}, {Barro}, {Bezanson}, {Franx}, {Lorenz}, {Ma}, {Mowla}, {Pasha}, {van Dokkum}, \& {Weisz}}]{Beverage_2024}
{Beverage}, A.~G., {Kriek}, M., {Suess}, K.~A., {et~al.} 2024, \apj, 966, 234

\bibitem[{{Beverage} {et~al.}(2025){Beverage}, {Slob}, {Kriek}, {Conroy}, {Barro}, {Bezanson}, {Brammer}, {Cheng}, {de Graaff}, {F{\"o}rster Schreiber}, {Franx}, {Lorenz}, {Mancera Pi{\~n}a}, {Marchesini}, {Muzzin}, {Newman}, {Price}, {Shapley}, {Stefanon}, {Suess}, {van Dokkum}, {Weinberg}, \& {Weisz}}]{Beverage_suspense}
{Beverage}, A.~G., {Slob}, M., {Kriek}, M., {et~al.} 2025, \apj, 979, 249

\bibitem[{{Bezanson} {et~al.}(2009){Bezanson}, {van Dokkum}, {Tal}, {Marchesini}, {Kriek}, {Franx}, \& {Coppi}}]{Bezanson_2009}
{Bezanson}, R., {van Dokkum}, P.~G., {Tal}, T., {et~al.} 2009, \apj, 697, 1290

\bibitem[{{Bluck} {et~al.}(2014){Bluck}, {Mendel}, {Ellison}, {Moreno}, {Simard}, {Patton}, \& {Starkenburg}}]{Bluck_2014}
{Bluck}, A. F.~L., {Mendel}, J.~T., {Ellison}, S.~L., {et~al.} 2014, \mnras, 441, 599

\bibitem[{Bradley {et~al.}(2024)Bradley, Sip{\H o}cz, Robitaille, Tollerud, Vin{\'{\i}}cius, Deil, Barbary, Wilson, Busko, Donath, G{\"u}nther, Cara, Lim, Me{\ss}linger, Burnett, Conseil, Droettboom, Bostroem, Bray, Bratholm, Jamieson, Ginsburg, Barentsen, Craig, Pascual, Rathi, Perrin, Morris, \& Perren}]{photutils}
Bradley, L., Sip{\H o}cz, B., Robitaille, T., {et~al.} 2024, astropy/photutils: 1.12.0

\bibitem[{Brammer(2023)}]{grizli}
Brammer, G. 2023, grizli

\bibitem[{{Bruzual} \& {Charlot}(2003)}]{Bruzual_2003}
{Bruzual}, G. \& {Charlot}, S. 2003, \mnras, 344, 1000

\bibitem[{Bushouse {et~al.}(2023)Bushouse, Eisenhamer, Dencheva, Davies, Greenfield, Morrison, Hodge, Simon, Grumm, Droettboom, Slavich, Sosey, Pauly, Miller, Jedrzejewski, Hack, Davis, Crawford, Law, Gordon, Regan, Cara, MacDonald, Bradley, Shanahan, Jamieson, Teodoro, Williams, \& Pena-Guerrero}]{Bushouse_2023}
Bushouse, H., Eisenhamer, J., Dencheva, N., {et~al.} 2023, JWST Calibration Pipeline

\bibitem[{{Carnall}(2017)}]{SPECTRES}
{Carnall}, A.~C. 2017, arXiv e-prints, arXiv:1705.05165

\bibitem[{{Carnall} {et~al.}(2022){Carnall}, {McLure}, {Dunlop}, {Hamadouche}, {Cullen}, {McLeod}, {Begley}, {Amorin}, {Bolzonella}, {Castellano}, {Cimatti}, {Fontanot}, {Gargiulo}, {Garilli}, {Mannucci}, {Pentericci}, {Talia}, {Zamorani}, {Calabro}, {Cresci}, \& {Hathi}}]{Carnall_2022}
{Carnall}, A.~C., {McLure}, R.~J., {Dunlop}, J.~S., {et~al.} 2022, \apj, 929, 131

\bibitem[{{Carollo} {et~al.}(2013){Carollo}, {Bschorr}, {Renzini}, {Lilly}, {Capak}, {Cibinel}, {Ilbert}, {Onodera}, {Scoville}, {Cameron}, {Mobasher}, {Sanders}, \& {Taniguchi}}]{Carollo_2013}
{Carollo}, C.~M., {Bschorr}, T.~J., {Renzini}, A., {et~al.} 2013, \apj, 773, 112

\bibitem[{{Casey} {et~al.}(2023){Casey}, {Kartaltepe}, {Drakos}, {Franco}, {Harish}, {Paquereau}, {Ilbert}, {Rose}, {Cox}, {Nightingale}, {Robertson}, {Silverman}, {Koekemoer}, {Massey}, {McCracken}, {Rhodes}, {Akins}, {Allen}, {Amvrosiadis}, {Arango-Toro}, {Bagley}, {Bongiorno}, {Capak}, {Champagne}, {Chartab}, {Ch{\'a}vez Ortiz}, {Chworowsky}, {Cooke}, {Cooper}, {Darvish}, {Ding}, {Faisst}, {Finkelstein}, {Fujimoto}, {Gentile}, {Gillman}, {Gould}, {Gozaliasl}, {Hayward}, {He}, {Hemmati}, {Hirschmann}, {Jahnke}, {Jin}, {Khostovan}, {Kokorev}, {Lambrides}, {Laigle}, {Larson}, {Leung}, {Liu}, {Liaudat}, {Long}, {Magdis}, {Mahler}, {Mainieri}, {Manning}, {Maraston}, {Martin}, {McCleary}, {McKinney}, {McPartland}, {Mobasher}, {Pattnaik}, {Renzini}, {Rich}, {Sanders}, {Sattari}, {Scognamiglio}, {Scoville}, {Sheth}, {Shuntov}, {Sparre}, {Suzuki}, {Talia}, {Toft}, {Trakhtenbrot}, {Urry}, {Valentino}, {Vanderhoof}, {Vardoulaki}, {Weaver}, {Whitaker}, {Wilkins}, {Yang}, \& {Zavala}}]{cosmos-web}
{Casey}, C.~M., {Kartaltepe}, J.~S., {Drakos}, N.~E., {et~al.} 2023, \apj, 954, 31

\bibitem[{{Cheng} {et~al.}(2024){Cheng}, {Kriek}, {Beverage}, {van der Wel}, {Bezanson}, {D'Eugenio}, {Franx}, {Mancera Pi{\~n}a}, {Nersesian}, {Slob}, {Suess}, {van Dokkum}, {Wu}, {Gallazzi}, \& {Zibetti}}]{Cheng_2024}
{Cheng}, C.~M., {Kriek}, M., {Beverage}, A.~G., {et~al.} 2024, \mnras, 532, 3604

\bibitem[{{Cheng} {et~al.}(2026){Cheng}, {Slob}, {Kriek}, {Beverage}, {van Dokkum}, {Bezanson}, {Brammer}, {Conroy}, {de Graaff}, {Eftekhari}, {Feldmann}, {Goesaert}, {Gu}, {Leja}, {Lorenz}, {Mancera Pi{\~n}a}, {Mart{\'\i}n-Navarro}, {Newman}, {Price}, {Shapley}, {Sharda}, {Suess}, {van der Wel}, \& {Weisz}}]{Cheng_2026}
{Cheng}, C.~M., {Slob}, M., {Kriek}, M., {et~al.} 2026, arXiv e-prints, arXiv:2601.20864

\bibitem[{{Choi} {et~al.}(2016){Choi}, {Dotter}, {Conroy}, {Cantiello}, {Paxton}, \& {Johnson}}]{MIST}
{Choi}, J., {Dotter}, A., {Conroy}, C., {et~al.} 2016, \apj, 823, 102

\bibitem[{{Ciocca} {et~al.}(2017){Ciocca}, {Saracco}, {Gargiulo}, \& {De Propris}}]{Ciocca_2017}
{Ciocca}, F., {Saracco}, P., {Gargiulo}, A., \& {De Propris}, R. 2017, \mnras, 466, 4492

\bibitem[{{Conroy} \& {van Dokkum}(2012)}]{CvD_2012a}
{Conroy}, C. \& {van Dokkum}, P. 2012, \apj, 747, 69

\bibitem[{{Conroy} {et~al.}(2018){Conroy}, {Villaume}, {van Dokkum}, \& {Lind}}]{Conroy_2018}
{Conroy}, C., {Villaume}, A., {van Dokkum}, P.~G., \& {Lind}, K. 2018, \apj, 854, 139

\bibitem[{{Croton} {et~al.}(2006){Croton}, {Springel}, {White}, {De Lucia}, {Frenk}, {Gao}, {Jenkins}, {Kauffmann}, {Navarro}, \& {Yoshida}}]{Croton_2006}
{Croton}, D.~J., {Springel}, V., {White}, S. D.~M., {et~al.} 2006, \mnras, 365, 11

\bibitem[{{Damjanov} {et~al.}(2023){Damjanov}, {Sohn}, {Geller}, {Utsumi}, \& {Dell'Antonio}}]{Damjanov_2023}
{Damjanov}, I., {Sohn}, J., {Geller}, M.~J., {Utsumi}, Y., \& {Dell'Antonio}, I. 2023, \apj, 943, 149

\bibitem[{{Damjanov} {et~al.}(2019){Damjanov}, {Zahid}, {Geller}, {Utsumi}, {Sohn}, \& {Souchereau}}]{Damjanov_2019}
{Damjanov}, I., {Zahid}, H.~J., {Geller}, M.~J., {et~al.} 2019, \apj, 872, 91

\bibitem[{{de Graaff} {et~al.}(2024){de Graaff}, {Rix}, {Carniani}, {Suess}, {Charlot}, {Curtis-Lake}, {Arribas}, {Baker}, {Boyett}, {Bunker}, {Cameron}, {Chevallard}, {Curti}, {Eisenstein}, {Franx}, {Hainline}, {Hausen}, {Ji}, {Johnson}, {Jones}, {Maiolino}, {Maseda}, {Nelson}, {Parlanti}, {Rawle}, {Robertson}, {Tacchella}, {{\"U}bler}, {Williams}, {Willmer}, \& {Willott}}]{msafit}
{de Graaff}, A., {Rix}, H.-W., {Carniani}, S., {et~al.} 2024, \aap, 684, A87

\bibitem[{{De Lucia} {et~al.}(2017){De Lucia}, {Fontanot}, \& {Hirschmann}}]{De_Lucia_2017}
{De Lucia}, G., {Fontanot}, F., \& {Hirschmann}, M. 2017, \mnras, 466, L88

\bibitem[{{Dekel} \& {Burkert}(2014)}]{Dekel_Burkert_2014}
{Dekel}, A. \& {Burkert}, A. 2014, \mnras, 438, 1870

\bibitem[{{D'Eugenio} {et~al.}(2024){D'Eugenio}, {P{\'e}rez-Gonz{\'a}lez}, {Maiolino}, {Scholtz}, {Perna}, {Circosta}, {{\"U}bler}, {Arribas}, {B{\"o}ker}, {Bunker}, {Carniani}, {Charlot}, {Chevallard}, {Cresci}, {Curtis-Lake}, {Jones}, {Kumari}, {Lamperti}, {Looser}, {Parlanti}, {Rix}, {Robertson}, {Rodr{\'\i}guez Del Pino}, {Tacchella}, {Venturi}, \& {Willott}}]{D'Eugenio_2024}
{D'Eugenio}, F., {P{\'e}rez-Gonz{\'a}lez}, P.~G., {Maiolino}, R., {et~al.} 2024, Nature Astronomy, 8, 1443

\bibitem[{{D'Eugenio} {et~al.}(2020){D'Eugenio}, {van der Wel}, {Wu}, {Barone}, {van Houdt}, {Bezanson}, {Straatman}, {Pacifici}, {Muzzin}, {Gallazzi}, {Wild}, {Sobral}, {Bell}, {Zibetti}, {Mowla}, \& {Franx}}]{D'Eugenio_2020}
{D'Eugenio}, F., {van der Wel}, A., {Wu}, P.-F., {et~al.} 2020, \mnras, 497, 389

\bibitem[{{Ditrani} {et~al.}(2022){Ditrani}, {Andreon}, {Longhetti}, \& {Newman}}]{Ditrani_2022}
{Ditrani}, F.~R., {Andreon}, S., {Longhetti}, M., \& {Newman}, A. 2022, \aap, 660, A132

\bibitem[{{Dressel} {et~al.}(2007){Dressel}, {Hodge}, \& {Barrett}}]{Dressel_2007}
{Dressel}, L., {Hodge}, P., \& {Barrett}, P. 2007, {wx2d: A PyRAF Routine to Resample Spectral Images}, Instrument Science Report STIS 2007-04, 20 pages

\bibitem[{{Dumont} {et~al.}(2025){Dumont}, {Neumayer}, {Seth}, {B{\"o}ker}, {Eracleous}, {Goold}, {Greene}, {G{\"u}ltekin}, {Ho}, {Walsh}, \& {L{\"u}tzgendorf}}]{wicked}
{Dumont}, A., {Neumayer}, N., {Seth}, A.~C., {et~al.} 2025, \aap, 703, A54

\bibitem[{{Ellison} {et~al.}(2018){Ellison}, {S{\'a}nchez}, {Ibarra-Medel}, {Antonio}, {Mendel}, \& {Barrera-Ballesteros}}]{Ellison_2018}
{Ellison}, S.~L., {S{\'a}nchez}, S.~F., {Ibarra-Medel}, H., {et~al.} 2018, \mnras, 474, 2039

\bibitem[{{Ferreras} {et~al.}(2019){Ferreras}, {Scott}, {La Barbera}, {Croom}, {van de Sande}, {Hopkins}, {Colless}, {Barone}, {d'Eugenio}, {Bland-Hawthorn}, {Brough}, {Bryant}, {Konstantopoulos}, {Lagos}, {Lawrence}, {L{\'o}pez-S{\'a}nchez}, {Medling}, {Owers}, \& {Richards}}]{Ferreras_2019}
{Ferreras}, I., {Scott}, N., {La Barbera}, F., {et~al.} 2019, \mnras, 489, 608

\bibitem[{{Ferruit} {et~al.}(2022){Ferruit}, {Jakobsen}, {Giardino}, {Rawle}, {Alves de Oliveira}, {Arribas}, {Beck}, {Birkmann}, {B{\"o}ker}, {Bunker}, {Charlot}, {de Marchi}, {Franx}, {Henry}, {Karakla}, {Kassin}, {Kumari}, {L{\'o}pez-Caniego}, {L{\"u}tzgendorf}, {Maiolino}, {Manjavacas}, {Marston}, {Moseley}, {Muzerolle}, {Pirzkal}, {Rauscher}, {Rix}, {Sabbi}, {Sirianni}, {te Plate}, {Valenti}, {Willott}, \& {Zeidler}}]{Ferruit_2022}
{Ferruit}, P., {Jakobsen}, P., {Giardino}, G., {et~al.} 2022, \aap, 661, A81

\bibitem[{{Fontanot} {et~al.}(2017){Fontanot}, {De Lucia}, {Hirschmann}, {Bruzual}, {Charlot}, \& {Zibetti}}]{Fontanot_2017}
{Fontanot}, F., {De Lucia}, G., {Hirschmann}, M., {et~al.} 2017, \mnras, 464, 3812

\bibitem[{{Foreman-Mackey} {et~al.}(2013){Foreman-Mackey}, {Hogg}, {Lang}, \& {Goodman}}]{emcee}
{Foreman-Mackey}, D., {Hogg}, D.~W., {Lang}, D., \& {Goodman}, J. 2013, \pasp, 125, 306

\bibitem[{{Franx} \& {Illingworth}(1990)}]{Franx_1990}
{Franx}, M. \& {Illingworth}, G. 1990, \apjl, 359, L41

\bibitem[{{Franx} {et~al.}(2008){Franx}, {van Dokkum}, {F{\"o}rster Schreiber}, {Wuyts}, {Labb{\'e}}, \& {Toft}}]{Franx_2008}
{Franx}, M., {van Dokkum}, P.~G., {F{\"o}rster Schreiber}, N.~M., {et~al.} 2008, \apj, 688, 770

\bibitem[{{Gallazzi} {et~al.}(2005){Gallazzi}, {Charlot}, {Brinchmann}, {White}, \& {Tremonti}}]{Gallazzi_2005}
{Gallazzi}, A., {Charlot}, S., {Brinchmann}, J., {White}, S. D.~M., \& {Tremonti}, C.~A. 2005, \mnras, 362, 41

\bibitem[{{Gallazzi} {et~al.}(2021){Gallazzi}, {Pasquali}, {Zibetti}, \& {Barbera}}]{Gallazzi_2021}
{Gallazzi}, A.~R., {Pasquali}, A., {Zibetti}, S., \& {Barbera}, F.~L. 2021, \mnras, 502, 4457

\bibitem[{{Gargiulo} {et~al.}(2012){Gargiulo}, {Saracco}, {Longhetti}, {La Barbera}, \& {Tamburri}}]{Gargiulo_2012}
{Gargiulo}, A., {Saracco}, P., {Longhetti}, M., {La Barbera}, F., \& {Tamburri}, S. 2012, \mnras, 425, 2698

\bibitem[{{Goddard} {et~al.}(2017){Goddard}, {Thomas}, {Maraston}, {Westfall}, {Etherington}, {Riffel}, {Mallmann}, {Zheng}, {Argudo-Fern{\'a}ndez}, {Bershady}, {Bundy}, {Drory}, {Law}, {Yan}, {Wake}, {Weijmans}, {Bizyaev}, {Brownstein}, {Lane}, {Maiolino}, {Masters}, {Merrifield}, {Nitschelm}, {Pan}, {Roman-Lopes}, \& {Storchi-Bergmann}}]{Goddard_2017}
{Goddard}, D., {Thomas}, D., {Maraston}, C., {et~al.} 2017, \mnras, 465, 688

\bibitem[{{Gonz{\'a}lez Delgado} {et~al.}(2015){Gonz{\'a}lez Delgado}, {Garc{\'\i}a-Benito}, {P{\'e}rez}, {Cid Fernandes}, {de Amorim}, {Cortijo-Ferrero}, {Lacerda}, {L{\'o}pez Fern{\'a}ndez}, {Vale-Asari}, {S{\'a}nchez}, {Moll{\'a}}, {Ruiz-Lara}, {S{\'a}nchez-Bl{\'a}zquez}, {Walcher}, {Alves}, {Aguerri}, {Bekerait{\'e}}, {Bland-Hawthorn}, {Galbany}, {Gallazzi}, {Husemann}, {Iglesias-P{\'a}ramo}, {Kalinova}, {L{\'o}pez-S{\'a}nchez}, {Marino}, {M{\'a}rquez}, {Masegosa}, {Mast}, {M{\'e}ndez-Abreu}, {Mendoza}, {del Olmo}, {P{\'e}rez}, {Quirrenbach}, \& {Zibetti}}]{Gonzalez_Delgado_2015}
{Gonz{\'a}lez Delgado}, R.~M., {Garc{\'\i}a-Benito}, R., {P{\'e}rez}, E., {et~al.} 2015, \aap, 581, A103

\bibitem[{{Greene} {et~al.}(2015){Greene}, {Janish}, {Ma}, {McConnell}, {Blakeslee}, {Thomas}, \& {Murphy}}]{Greene_2015}
{Greene}, J.~E., {Janish}, R., {Ma}, C.-P., {et~al.} 2015, \apj, 807, 11

\bibitem[{{Greene} {et~al.}(2013){Greene}, {Murphy}, {Graves}, {Gunn}, {Raskutti}, {Comerford}, \& {Gebhardt}}]{Greene_2013}
{Greene}, J.~E., {Murphy}, J.~D., {Graves}, G.~J., {et~al.} 2013, \apj, 776, 64

\bibitem[{Greene {et~al.}(2019)Greene, Veale, Ma, Thomas, Quenneville, Blakeslee, Walsh, Goulding, \& Ito}]{Greene_2019}
Greene, J.~E., Veale, M., Ma, C.-P., {et~al.} 2019, \apj, 874, 66

\bibitem[{{Griffith} {et~al.}(2012){Griffith}, {Cooper}, {Newman}, {Moustakas}, {Stern}, {Comerford}, {Davis}, {Lotz}, {Barden}, {Conselice}, {Capak}, {Faber}, {Kirkpatrick}, {Koekemoer}, {Koo}, {Noeske}, {Scoville}, {Sheth}, {Shopbell}, {Willmer}, \& {Weiner}}]{Griffith_2012}
{Griffith}, R.~L., {Cooper}, M.~C., {Newman}, J.~A., {et~al.} 2012, \apjs, 200, 9

\bibitem[{{Gu} {et~al.}(2022){Gu}, {Greene}, {Newman}, {Kreisch}, {Quenneville}, {Ma}, \& {Blakeslee}}]{Gu_2022}
{Gu}, M., {Greene}, J.~E., {Newman}, A.~B., {et~al.} 2022, \apj, 932, 103

\bibitem[{{Guo} {et~al.}(2011){Guo}, {Giavalisco}, {Cassata}, {Ferguson}, {Dickinson}, {Renzini}, {Koekemoer}, {Grogin}, {Papovich}, {Tundo}, {Fontana}, {Lotz}, \& {Salimbeni}}]{Guo_2011}
{Guo}, Y., {Giavalisco}, M., {Cassata}, P., {et~al.} 2011, \apj, 735, 18

\bibitem[{{Haryana} {et~al.}(2025){Haryana}, {Akiyama}, {Abdurro'uf}, {Wulandari}, {Alfonzo}, {Lee}, {Matsumoto}, {Sutanto}, {Effendi}, {Fitriana}, {Huda}, {Jaelani}, {Kusuma}, {Puspitarini}, \& {Triani}}]{Saputra_2025}
{Haryana}, N.~S., {Akiyama}, M., {Abdurro'uf}, {et~al.} 2025, \apj, 994, 215

\bibitem[{{Hopkins} {et~al.}(2009){Hopkins}, {Cox}, {Dutta}, {Hernquist}, {Kormendy}, \& {Lauer}}]{Hopkins_2009}
{Hopkins}, P.~F., {Cox}, T.~J., {Dutta}, S.~N., {et~al.} 2009, \apjs, 181, 135

\bibitem[{{Hopkins} {et~al.}(2008){Hopkins}, {Cox}, {Kere{\v{s}}}, \& {Hernquist}}]{Hopkins_2008}
{Hopkins}, P.~F., {Cox}, T.~J., {Kere{\v{s}}}, D., \& {Hernquist}, L. 2008, \apjs, 175, 390

\bibitem[{{Hopkins} {et~al.}(2010){Hopkins}, {Croton}, {Bundy}, {Khochfar}, {van den Bosch}, {Somerville}, {Wetzel}, {Keres}, {Hernquist}, {Stewart}, {Younger}, {Genel}, \& {Ma}}]{Hopkins_2010}
{Hopkins}, P.~F., {Croton}, D., {Bundy}, K., {et~al.} 2010, \apj, 724, 915

\bibitem[{{Jafariyazani} {et~al.}(2025){Jafariyazani}, {Newman}, {Mobasher}, {Belli}, {Ellis}, \& {Faisst}}]{Jafariyazani_2024}
{Jafariyazani}, M., {Newman}, A.~B., {Mobasher}, B., {et~al.} 2025, \apj, 986, 148

\bibitem[{{Jafariyazani} {et~al.}(2020){Jafariyazani}, {Newman}, {Mobasher}, {Belli}, {Ellis}, \& {Patel}}]{Jafariyazani_2020}
{Jafariyazani}, M., {Newman}, A.~B., {Mobasher}, B., {et~al.} 2020, \apjl, 897, L42

\bibitem[{{Jones} {et~al.}(2015){Jones}, {Wang}, {Schmidt}, {Treu}, {Brammer}, {Brada{\v{c}}}, {Dressler}, {Henry}, {Malkan}, {Pentericci}, \& {Trenti}}]{Jones_2015}
{Jones}, T., {Wang}, X., {Schmidt}, K.~B., {et~al.} 2015, \aj, 149, 107

\bibitem[{{Ju} {et~al.}(2025){Ju}, {Wang}, {Jones}, {Bari{\v{s}}i{\'c}}, {Nanayakkara}, {Bundy}, {Faucher-Gigu{\`e}re}, {Feng}, {Glazebrook}, {Henry}, {Malkan}, {Obreschkow}, {Roy}, {Sanders}, {Sun}, {Treu}, \& {Zhou}}]{Ju_2025}
{Ju}, M., {Wang}, X., {Jones}, T., {et~al.} 2025, \apjl, 978, L39

\bibitem[{{Kauffmann} {et~al.}(2003){Kauffmann}, {Heckman}, {White}, {Charlot}, {Tremonti}, {Peng}, {Seibert}, {Brinkmann}, {Nichol}, {SubbaRao}, \& {York}}]{Kauffmann_2003}
{Kauffmann}, G., {Heckman}, T.~M., {White}, S. D.~M., {et~al.} 2003, \mnras, 341, 54

\bibitem[{{Keating} {et~al.}(2015){Keating}, {Abraham}, {Schiavon}, {Graves}, {Damjanov}, {Yan}, {Newman}, \& {Simard}}]{Keating_2015}
{Keating}, S.~K., {Abraham}, R.~G., {Schiavon}, R., {et~al.} 2015, \apj, 798, 26

\bibitem[{{Koekemoer} {et~al.}(2007){Koekemoer}, {Aussel}, {Calzetti}, {Capak}, {Giavalisco}, {Kneib}, {Leauthaud}, {Le F{\`e}vre}, {McCracken}, {Massey}, {Mobasher}, {Rhodes}, {Scoville}, \& {Shopbell}}]{COSMOS_ACS_mosaics}
{Koekemoer}, A.~M., {Aussel}, H., {Calzetti}, D., {et~al.} 2007, \apjs, 172, 196

\bibitem[{{Kriek} {et~al.}(2016){Kriek}, {Conroy}, {van Dokkum}, {Shapley}, {Choi}, {Reddy}, {Siana}, {van de Voort}, {Coil}, \& {Mobasher}}]{Kriek_2016}
{Kriek}, M., {Conroy}, C., {van Dokkum}, P.~G., {et~al.} 2016, \nat, 540, 248

\bibitem[{Kriek {et~al.}(2019)Kriek, Price, Conroy, Suess, Mowla, Pasha, Bezanson, van Dokkum, \& Barro}]{Kriek_2019}
Kriek, M., Price, S.~H., Conroy, C., {et~al.} 2019, \apj, 880, L31

\bibitem[{{Kroupa}(2001)}]{Kroupa}
{Kroupa}, P. 2001, \mnras, 322, 231

\bibitem[{{Kuntschner} {et~al.}(2006){Kuntschner}, {Emsellem}, {Bacon}, {Bureau}, {Cappellari}, {Davies}, {de Zeeuw}, {Falc{\'o}n-Barroso}, {Krajnovi{\'c}}, {McDermid}, {Peletier}, \& {Sarzi}}]{Kuntschner_2006}
{Kuntschner}, H., {Emsellem}, E., {Bacon}, R., {et~al.} 2006, \mnras, 369, 497

\bibitem[{{Kuntschner} {et~al.}(2010){Kuntschner}, {Emsellem}, {Bacon}, {Cappellari}, {Davies}, {de Zeeuw}, {Falc{\'o}n-Barroso}, {Krajnovi{\'c}}, {McDermid}, {Peletier}, {Sarzi}, {Shapiro}, {van den Bosch}, \& {van de Ven}}]{Kuntschner_2010}
{Kuntschner}, H., {Emsellem}, E., {Bacon}, R., {et~al.} 2010, \mnras, 408, 97

\bibitem[{{La Barbera} {et~al.}(2005){La Barbera}, {de Carvalho}, {Gal}, {Busarello}, {Merluzzi}, {Capaccioli}, \& {Djorgovski}}]{La_Barbera_2005}
{La Barbera}, F., {de Carvalho}, R.~R., {Gal}, R.~R., {et~al.} 2005, \apjl, 626, L19

\bibitem[{{Law} {et~al.}(2023){Law}, {E. Morrison}, {Argyriou}, {Patapis}, {{\'A}lvarez-M{\'a}rquez}, {Labiano}, \& {Vandenbussche}}]{Law_2023}
{Law}, D.~R., {E. Morrison}, J., {Argyriou}, I., {et~al.} 2023, \aj, 166, 45

\bibitem[{{Leja} {et~al.}(2019){Leja}, {Carnall}, {Johnson}, {Conroy}, \& {Speagle}}]{Leja_2019_sfh}
{Leja}, J., {Carnall}, A.~C., {Johnson}, B.~D., {Conroy}, C., \& {Speagle}, J.~S. 2019, \apj, 876, 3

\bibitem[{{Li} {et~al.}(2018){Li}, {Mao}, {Cappellari}, {Ge}, {Long}, {Li}, {Mo}, {Li}, {Zheng}, {Bundy}, {Thomas}, {Brownstein}, {Roman Lopes}, {Law}, \& {Drory}}]{Li_2018}
{Li}, H., {Mao}, S., {Cappellari}, M., {et~al.} 2018, \mnras, 476, 1765

\bibitem[{{Lupton} {et~al.}(2004){Lupton}, {Blanton}, {Fekete}, {Hogg}, {O'Mullane}, {Szalay}, \& {Wherry}}]{Lupton_2004}
{Lupton}, R., {Blanton}, M.~R., {Fekete}, G., {et~al.} 2004, \pasp, 116, 133

\bibitem[{{Maiolino} \& {Mannucci}(2019)}]{Maiolino_Mannucci_2019}
{Maiolino}, R. \& {Mannucci}, F. 2019, \aapr, 27, 3

\bibitem[{{Mann} {et~al.}(2015){Mann}, {Feiden}, {Gaidos}, {Boyajian}, \& {von Braun}}]{Mann_2015}
{Mann}, A.~W., {Feiden}, G.~A., {Gaidos}, E., {Boyajian}, T., \& {von Braun}, K. 2015, \apj, 804, 64

\bibitem[{{Mart{\'\i}n-Navarro}(2016)}]{Martin_Navarro_2016}
{Mart{\'\i}n-Navarro}, I. 2016, \mnras, 456, L104

\bibitem[{{Mart{\'\i}n-Navarro} {et~al.}(2018){Mart{\'\i}n-Navarro}, {Vazdekis}, {Falc{\'o}n-Barroso}, {La Barbera}, {Y{\i}ld{\i}r{\i}m}, \& {van de Ven}}]{Martin_Navarro_2018}
{Mart{\'\i}n-Navarro}, I., {Vazdekis}, A., {Falc{\'o}n-Barroso}, J., {et~al.} 2018, \mnras, 475, 3700

\bibitem[{{Martorano} {et~al.}(2026){Martorano}, {van der Wel}, {Gebek}, {Baes}, {Bell}, {Brammer}, {Meidt}, {Nersesian}, {Whitaker}, \& {Wuyts}}]{Martorano_2026}
{Martorano}, M., {van der Wel}, A., {Gebek}, A., {et~al.} 2026, \aap, 705, A236

\bibitem[{{Maseda} {et~al.}(2023){Maseda}, {Lewis}, {Matthee}, {Hennawi}, {Boogaard}, {Feltre}, {Nanayakkara}, {Bacon}, {Barger}, {Brinchmann}, {Franx}, {Hashimoto}, {Inami}, {Kusakabe}, {Leclercq}, {Rowland}, {Taylor}, {Tremonti}, {Urrutia}, {Schaye}, {Simmonds}, \& {Vitte}}]{Maseda_2023}
{Maseda}, M.~V., {Lewis}, Z., {Matthee}, J., {et~al.} 2023, \apj, 956, 11

\bibitem[{{Matteucci}(1994)}]{Matteucci_1994}
{Matteucci}, F. 1994, \aap, 288, 57

\bibitem[{{McCracken} {et~al.}(2012){McCracken}, {Milvang-Jensen}, {Dunlop}, {Franx}, {Fynbo}, {Le F{\`e}vre}, {Holt}, {Caputi}, {Goranova}, {Buitrago}, {Emerson}, {Freudling}, {Hudelot}, {L{\'o}pez-Sanjuan}, {Magnard}, {Mellier}, {M{\o}ller}, {Nilsson}, {Sutherland}, {Tasca}, \& {Zabl}}]{McCracken_2012}
{McCracken}, H.~J., {Milvang-Jensen}, B., {Dunlop}, J., {et~al.} 2012, \aap, 544, A156

\bibitem[{{McGrath} {et~al.}(2026){McGrath}, {Finkelstein}, {Barro}, {Pandya}, {Ferguson}, {Kartaltepe}, {Kocevski}, {Amor{\'\i}n}, {Backhaus}, {Buitrago}, {Calabr{\`o}}, {Cheng}, {Costantin}, {Cox}, {Davis}, {Gandolfi}, {Guo}, {Hathi}, {Hirschmann}, {Holwerda}, {Huertas-Company}, {Koekemoer}, {Lucas}, {Mobasher}, {Pacucci}, {Papovich}, {P{\'e}rez-Gonz{\'a}lez}, {Trump}, {Yung}, {Arrabal Haro}, {Bagley}, {Dickinson}, {Fontana}, {Grazian}, {Grogin}, {Kewley}, {Kirkpatrick}, {Lotz}, {Pentericci}, {Pirzkal}, {Ravindranath}, {Somerville}, {Wilkins}, {Yang}, {Seill{\'e}}, \& {Wang}}]{Mcgrath_2026}
{McGrath}, E.~J., {Finkelstein}, S.~L., {Barro}, G., {et~al.} 2026, \apjl, 999, L6

\bibitem[{{Mehlert} {et~al.}(2003){Mehlert}, {Thomas}, {Saglia}, {Bender}, \& {Wegner}}]{Mehlert_2003}
{Mehlert}, D., {Thomas}, D., {Saglia}, R.~P., {Bender}, R., \& {Wegner}, G. 2003, \aap, 407, 423

\bibitem[{{Miller} {et~al.}(2023){Miller}, {van Dokkum}, \& {Mowla}}]{Miller_2023}
{Miller}, T.~B., {van Dokkum}, P., \& {Mowla}, L. 2023, \apj, 945, 155

\bibitem[{{Miller} {et~al.}(2022){Miller}, {Whitaker}, {Nelson}, {van Dokkum}, {Bezanson}, {Brammer}, {Heintz}, {Leja}, {Suess}, \& {Weaver}}]{Miller_2022}
{Miller}, T.~B., {Whitaker}, K.~E., {Nelson}, E.~J., {et~al.} 2022, \apjl, 941, L37

\bibitem[{{Moffat}(1969)}]{Moffat_1969}
{Moffat}, A.~F.~J. 1969, \aap, 3, 455

\bibitem[{{Muzzin} {et~al.}(2013{\natexlab{a}}){Muzzin}, {Marchesini}, {Stefanon}, {Franx}, {McCracken}, {Milvang-Jensen}, {Dunlop}, {Fynbo}, {Brammer}, {Labb{\'e}}, \& {van Dokkum}}]{Muzzin_2013_uvj}
{Muzzin}, A., {Marchesini}, D., {Stefanon}, M., {et~al.} 2013{\natexlab{a}}, \apj, 777, 18

\bibitem[{{Muzzin} {et~al.}(2013{\natexlab{b}}){Muzzin}, {Marchesini}, {Stefanon}, {Franx}, {Milvang-Jensen}, {Dunlop}, {Fynbo}, {Brammer}, {Labb{\'e}}, \& {van Dokkum}}]{Muzzin_2013}
{Muzzin}, A., {Marchesini}, D., {Stefanon}, M., {et~al.} 2013{\natexlab{b}}, \apjs, 206, 8

\bibitem[{{Naab} {et~al.}(2009){Naab}, {Johansson}, \& {Ostriker}}]{Naab_2009}
{Naab}, T., {Johansson}, P.~H., \& {Ostriker}, J.~P. 2009, \apjl, 699, L178

\bibitem[{{Newman} {et~al.}(2012){Newman}, {Ellis}, {Bundy}, \& {Treu}}]{Newman_2012}
{Newman}, A.~B., {Ellis}, R.~S., {Bundy}, K., \& {Treu}, T. 2012, \apj, 746, 162

\bibitem[{{Newman} {et~al.}(2025){Newman}, {Gu}, {Belli}, {Ellis}, {Gangula}, {Greene}, {Walsh}, {Suyu}, {Ertl}, {Caminha}, {Granata}, {Grillo}, {Schuldt}, {Barone}, {Bird}, {Glazebrook}, {Jafariyazani}, {Kriek}, {Matthews}, {Morishita}, {Nanayakkara}, {Pierel}, {Acebr\textbackslash'on}, {Bergamini}, {Cha}, {Diego}, {Foo}, {Frye}, {Fudamoto}, {Jee}, {Kamieneski}, {Koekemoer}, {Meena}, {Nishida}, {Oguri}, {Rosati}, \& {Zitrin}}]{Newman_2025}
{Newman}, A.~B., {Gu}, M., {Belli}, S., {et~al.} 2025, arXiv e-prints, arXiv:2503.17478

\bibitem[{{Oke} \& {Gunn}(1983)}]{Oke_1983}
{Oke}, J.~B. \& {Gunn}, J.~E. 1983, \apj, 266, 713

\bibitem[{{Oser} {et~al.}(2012){Oser}, {Naab}, {Ostriker}, \& {Johansson}}]{Oser_2012}
{Oser}, L., {Naab}, T., {Ostriker}, J.~P., \& {Johansson}, P.~H. 2012, \apj, 744, 63

\bibitem[{{Oser} {et~al.}(2010){Oser}, {Ostriker}, {Naab}, {Johansson}, \& {Burkert}}]{Oser_2010}
{Oser}, L., {Ostriker}, J.~P., {Naab}, T., {Johansson}, P.~H., \& {Burkert}, A. 2010, \apj, 725, 2312

\bibitem[{{Oyarz{\'u}n} {et~al.}(2019){Oyarz{\'u}n}, {Bundy}, {Westfall}, {Belfiore}, {Thomas}, {Maraston}, {Lian}, {Arag{\'o}n-Salamanca}, {Zheng}, {Gonzalez-Perez}, {Law}, {Drory}, \& {Andrews}}]{Oyarzun_2019}
{Oyarz{\'u}n}, G.~A., {Bundy}, K., {Westfall}, K.~B., {et~al.} 2019, \apj, 880, 111

\bibitem[{{Oyarz{\'u}n} {et~al.}(2023){Oyarz{\'u}n}, {Bundy}, {Westfall}, {Lacerna}, {Yan}, {Brownstein}, {Drory}, \& {Lane}}]{Oyarzun_2023}
{Oyarz{\'u}n}, G.~A., {Bundy}, K., {Westfall}, K.~B., {et~al.} 2023, \apj, 947, 13

\bibitem[{{Parikh} {et~al.}(2024){Parikh}, {Saglia}, {Thomas}, {Mehrgan}, {Bender}, \& {Maraston}}]{Parikh_2024}
{Parikh}, T., {Saglia}, R., {Thomas}, J., {et~al.} 2024, \mnras, 528, 7338

\bibitem[{{Park} {et~al.}(2024){Park}, {Belli}, {Conroy}, {Johnson}, {Davies}, {Leja}, {Tacchella}, {Mendel}, {Benton}, {Bugiani}, {Emami}, {Khoram}, {Li}, {Maheson}, {Mathews}, {Naidu}, {Nelson}, {Terrazas}, \& {Weinberger}}]{Park_2024}
{Park}, M., {Belli}, S., {Conroy}, C., {et~al.} 2024, \apj, 976, 72

\bibitem[{{Pasquali} {et~al.}(2010){Pasquali}, {Gallazzi}, {Fontanot}, {van den Bosch}, {De Lucia}, {Mo}, \& {Yang}}]{Pasquali_2010}
{Pasquali}, A., {Gallazzi}, A., {Fontanot}, F., {et~al.} 2010, \mnras, 407, 937

\bibitem[{{Pastorello} {et~al.}(2014){Pastorello}, {Forbes}, {Foster}, {Brodie}, {Usher}, {Romanowsky}, {Strader}, \& {Arnold}}]{Pastorello_2014}
{Pastorello}, N., {Forbes}, D.~A., {Foster}, C., {et~al.} 2014, \mnras, 442, 1003

\bibitem[{{Pathak} {et~al.}(2021){Pathak}, {Belli}, \& {Weinberger}}]{Pathak_2021}
{Pathak}, D., {Belli}, S., \& {Weinberger}, R. 2021, \apjl, 916, L23

\bibitem[{{Peletier} \& {Valentijn}(1989)}]{Peletier_1989}
{Peletier}, R.~F. \& {Valentijn}, E.~A. 1989, \apss, 156, 127

\bibitem[{{Peng} {et~al.}(2010){Peng}, {Ho}, {Impey}, \& {Rix}}]{galfit}
{Peng}, C.~Y., {Ho}, L.~C., {Impey}, C.~D., \& {Rix}, H.-W. 2010, \aj, 139, 2097

\bibitem[{{P{\'e}rez-Gonz{\'a}lez} {et~al.}(2025){P{\'e}rez-Gonz{\'a}lez}, {D'Eugenio}, {Rodr{\'\i}guez del Pino}, {Perna}, {{\"U}bler}, {Maiolino}, {Arribas}, {Cresci}, {Lamperti}, {Bunker}, {Carniani}, {Charlot}, {Willott}, {B{\"o}ker}, {Parlanti}, {Scholtz}, {Venturi}, {Barro}, {Costantin}, {Mart{\'\i}n-Navarro}, {Dunlop}, \& {Magee}}]{Perez_Gonzalez_2024}
{P{\'e}rez-Gonz{\'a}lez}, P.~G., {D'Eugenio}, F., {Rodr{\'\i}guez del Pino}, B., {et~al.} 2025, Nature Astronomy, 9, 1240

\bibitem[{{Perna} {et~al.}(2023){Perna}, {Arribas}, {Marshall}, {D'Eugenio}, {{\"U}bler}, {Bunker}, {Charlot}, {Carniani}, {Jakobsen}, {Maiolino}, {Rodr{\'\i}guez Del Pino}, {Willott}, {B{\"o}ker}, {Circosta}, {Cresci}, {Curti}, {Husemann}, {Kumari}, {Lamperti}, {P{\'e}rez-Gonz{\'a}lez}, \& {Scholtz}}]{Perna_2023}
{Perna}, M., {Arribas}, S., {Marshall}, M., {et~al.} 2023, \aap, 679, A89

\bibitem[{{Perrin} {et~al.}(2015){Perrin}, {Long}, {Sivaramakrishnan}, {Lajoie}, {Elliot}, {Pueyo}, \& {Albert}}]{stpsf}
{Perrin}, M.~D., {Long}, J., {Sivaramakrishnan}, A., {et~al.} 2015, {WebbPSF: James Webb Space Telescope PSF Simulation Tool}, Astrophysics Source Code Library, record ascl:1504.007

\bibitem[{{Poggianti} {et~al.}(2013){Poggianti}, {Moretti}, {Calvi}, {D'Onofrio}, {Valentinuzzi}, {Fritz}, \& {Renzini}}]{Poggianti_2013}
{Poggianti}, B.~M., {Moretti}, A., {Calvi}, R., {et~al.} 2013, \apj, 777, 125

\bibitem[{{Price} {et~al.}(2016){Price}, {Kriek}, {Shapley}, {Reddy}, {Freeman}, {Coil}, {de Groot}, {Shivaei}, {Siana}, {Azadi}, {Barro}, {Mobasher}, {Sanders}, \& {Zick}}]{Price_2016}
{Price}, S.~H., {Kriek}, M., {Shapley}, A.~E., {et~al.} 2016, \apj, 819, 80

\bibitem[{{Quilis} \& {Trujillo}(2013)}]{Quilis_2013}
{Quilis}, V. \& {Trujillo}, I. 2013, \apjl, 773, L8

\bibitem[{{Rodriguez-Gomez} {et~al.}(2016){Rodriguez-Gomez}, {Pillepich}, {Sales}, {Genel}, {Vogelsberger}, {Zhu}, {Wellons}, {Nelson}, {Torrey}, {Springel}, {Ma}, \& {Hernquist}}]{Rodriguez_Gomez_2016}
{Rodriguez-Gomez}, V., {Pillepich}, A., {Sales}, L.~V., {et~al.} 2016, \mnras, 458, 2371

\bibitem[{{S{\'a}nchez-Bl{\'a}zquez} {et~al.}(2007){S{\'a}nchez-Bl{\'a}zquez}, {Forbes}, {Strader}, {Brodie}, \& {Proctor}}]{Sanchez_Blazquez_2007}
{S{\'a}nchez-Bl{\'a}zquez}, P., {Forbes}, D.~A., {Strader}, J., {Brodie}, J., \& {Proctor}, R. 2007, \mnras, 377, 759

\bibitem[{{S{\'a}nchez-Bl{\'a}zquez} {et~al.}(2006){S{\'a}nchez-Bl{\'a}zquez}, {Peletier}, {Jim{\'e}nez-Vicente}, {Cardiel}, {Cenarro}, {Falc{\'o}n-Barroso}, {Gorgas}, {Selam}, \& {Vazdekis}}]{Sanchez_Blazquez_2006}
{S{\'a}nchez-Bl{\'a}zquez}, P., {Peletier}, R.~F., {Jim{\'e}nez-Vicente}, J., {et~al.} 2006, \mnras, 371, 703

\bibitem[{{Santucci} {et~al.}(2020){Santucci}, {Brough}, {Scott}, {Montes}, {Owers}, {van Sande}, {Bland-Hawthorn}, {Bryant}, {Croom}, {Ferreras}, {Lawrence}, {L{\'o}pez-S{\'a}nchez}, \& {Richards}}]{Santucci_2020}
{Santucci}, G., {Brough}, S., {Scott}, N., {et~al.} 2020, \apj, 896, 75

\bibitem[{{Scoville} {et~al.}(2007){Scoville}, {Aussel}, {Brusa}, {Capak}, {Carollo}, {Elvis}, {Giavalisco}, {Guzzo}, {Hasinger}, {Impey}, {Kneib}, {LeFevre}, {Lilly}, {Mobasher}, {Renzini}, {Rich}, {Sanders}, {Schinnerer}, {Schminovich}, {Shopbell}, {Taniguchi}, \& {Tyson}}]{Scoville_2007}
{Scoville}, N., {Aussel}, H., {Brusa}, M., {et~al.} 2007, \apjs, 172, 1

\bibitem[{{Setton} {et~al.}(2024){Setton}, {Khullar}, {Miller}, {Bezanson}, {Greene}, {Suess}, {Whitaker}, {Antwi-Danso}, {Atek}, {Brammer}, {Cutler}, {Dayal}, {Feldmann}, {Fujimoto}, {Furtak}, {Glazebrook}, {Goulding}, {Kokorev}, {Labbe}, {Leja}, {Ma}, {Marchesini}, {Nanayakkara}, {Pan}, {Price}, {Siegel}, {Shipley}, {Weaver}, {van Dokkum}, {Wang}, \& {Williams}}]{Setton_2024}
{Setton}, D.~J., {Khullar}, G., {Miller}, T.~B., {et~al.} 2024, \apj, 974, 145

\bibitem[{{Shen} {et~al.}(2024){Shen}, {Papovich}, {Matharu}, {Pirzkal}, {Hu}, {Backhaus}, {Bagley}, {Cheng}, {Cleri}, {Finkelstein}, {Huertas-Company}, {Giavalisco}, {Grogin}, {Jung}, {Kartaltepe}, {Koekemoer}, {Lotz}, {Maseda}, {P{\'e}rez-Gonz{\'a}lez}, {Rothberg}, {Simons}, {Tacchella}, {Williams}, \& {Yung}}]{Shen_2024}
{Shen}, L., {Papovich}, C., {Matharu}, J., {et~al.} 2024, \apjl, 963, L49

\bibitem[{{Slob} {et~al.}(2024){Slob}, {Kriek}, {Beverage}, {Suess}, {Barro}, {Bezanson}, {Brammer}, {Cheng}, {Conroy}, {de Graaff}, {F{\"o}rster Schreiber}, {Franx}, {Lorenz}, {Mancera Pi{\~n}a}, {Marchesini}, {Muzzin}, {Newman}, {Price}, {Shapley}, {Stefanon}, {van Dokkum}, \& {Weisz}}]{Slob_2024}
{Slob}, M., {Kriek}, M., {Beverage}, A.~G., {et~al.} 2024, \apj, 973, 131

\bibitem[{{Slob} {et~al.}(2025){Slob}, {Kriek}, {de Graaff}, {Cheng}, {Beverage}, {Bezanson}, {F{\"o}rster Schreiber}, {Lorenz}, {Mancera Pi{\~n}a}, {Marchesini}, {Muzzin}, {Newman}, {Price}, {Suess}, {van de Sande}, {van Dokkum}, \& {Weisz}}]{Slob_2025}
{Slob}, M., {Kriek}, M., {de Graaff}, A., {et~al.} 2025, \aap, 702, A110

\bibitem[{{Smith} {et~al.}(2007){Smith}, {Armus}, {Dale}, {Roussel}, {Sheth}, {Buckalew}, {Jarrett}, {Helou}, \& {Kennicutt}}]{Smith_2007}
{Smith}, J.~D.~T., {Armus}, L., {Dale}, D.~A., {et~al.} 2007, \pasp, 119, 1133

\bibitem[{{Snyder} {et~al.}(2011){Snyder}, {Cox}, {Hayward}, {Hernquist}, \& {Jonsson}}]{Snyder_2011}
{Snyder}, G.~F., {Cox}, T.~J., {Hayward}, C.~C., {Hernquist}, L., \& {Jonsson}, P. 2011, \apj, 741, 77

\bibitem[{{Speagle}(2020)}]{dynesty}
{Speagle}, J.~S. 2020, \mnras, 493, 3132

\bibitem[{{Spilker} {et~al.}(2019){Spilker}, {Bezanson}, {Weiner}, {Whitaker}, \& {Williams}}]{Spilker_2019}
{Spilker}, J.~S., {Bezanson}, R., {Weiner}, B.~J., {Whitaker}, K.~E., \& {Williams}, C.~C. 2019, \apj, 883, 81

\bibitem[{{Spitoni} {et~al.}(2017){Spitoni}, {Vincenzo}, \& {Matteucci}}]{Spitoni_2017}
{Spitoni}, E., {Vincenzo}, F., \& {Matteucci}, F. 2017, \aap, 599, A6

\bibitem[{{Spolaor} {et~al.}(2010){Spolaor}, {Kobayashi}, {Forbes}, {Couch}, \& {Hau}}]{Spolaor_2010}
{Spolaor}, M., {Kobayashi}, C., {Forbes}, D.~A., {Couch}, W.~J., \& {Hau}, G. K.~T. 2010, \mnras, 408, 272

\bibitem[{{Stringer} {et~al.}(2015){Stringer}, {Trujillo}, {Dalla Vecchia}, \& {Martinez-Valpuesta}}]{Stringer_2015}
{Stringer}, M., {Trujillo}, I., {Dalla Vecchia}, C., \& {Martinez-Valpuesta}, I. 2015, \mnras, 449, 2396

\bibitem[{{Suess} {et~al.}(2022){Suess}, {Bezanson}, {Nelson}, {Setton}, {Price}, {van Dokkum}, {Brammer}, {Labb{\'e}}, {Leja}, {Miller}, {Robertson}, {Wel}, {Weaver}, \& {Whitaker}}]{Suess_2022}
{Suess}, K.~A., {Bezanson}, R., {Nelson}, E.~J., {et~al.} 2022, \apjl, 937, L33

\bibitem[{{Suess} {et~al.}(2019{\natexlab{a}}){Suess}, {Kriek}, {Price}, \& {Barro}}]{Suess_2019a}
{Suess}, K.~A., {Kriek}, M., {Price}, S.~H., \& {Barro}, G. 2019{\natexlab{a}}, \apj, 877, 103

\bibitem[{{Suess} {et~al.}(2019{\natexlab{b}}){Suess}, {Kriek}, {Price}, \& {Barro}}]{Suess_2019b}
{Suess}, K.~A., {Kriek}, M., {Price}, S.~H., \& {Barro}, G. 2019{\natexlab{b}}, \apjl, 885, L22

\bibitem[{{Suess} {et~al.}(2020){Suess}, {Kriek}, {Price}, \& {Barro}}]{Suess_2020}
{Suess}, K.~A., {Kriek}, M., {Price}, S.~H., \& {Barro}, G. 2020, \apjl, 899, L26

\bibitem[{{Suess} {et~al.}(2021){Suess}, {Kriek}, {Price}, \& {Barro}}]{Suess_2021}
{Suess}, K.~A., {Kriek}, M., {Price}, S.~H., \& {Barro}, G. 2021, \apj, 915, 87

\bibitem[{{Suess} {et~al.}(2023){Suess}, {Williams}, {Robertson}, {Ji}, {Johnson}, {Nelson}, {Alberts}, {Hainline}, {D'Eugenio}, {{\"U}bler}, {Rieke}, {Rieke}, {Bunker}, {Carniani}, {Charlot}, {Eisenstein}, {Maiolino}, {Stark}, {Tacchella}, \& {Willott}}]{Suess_2023}
{Suess}, K.~A., {Williams}, C.~C., {Robertson}, B., {et~al.} 2023, \apjl, 956, L42

\bibitem[{{Sybilska} {et~al.}(2018){Sybilska}, {Kuntschner}, {van de Ven}, {Vazdekis}, {Falc{\'o}n-Barroso}, {Peletier}, \& {Lisker}}]{Sybilska_2018}
{Sybilska}, A., {Kuntschner}, H., {van de Ven}, G., {et~al.} 2018, \mnras, 476, 4501

\bibitem[{{Szomoru} {et~al.}(2013){Szomoru}, {Franx}, {van Dokkum}, {Trenti}, {Illingworth}, {Labb{\'e}}, \& {Oesch}}]{Szomoru_2013}
{Szomoru}, D., {Franx}, M., {van Dokkum}, P.~G., {et~al.} 2013, \apj, 763, 73

\bibitem[{{Tacchella} {et~al.}(2018){Tacchella}, {Carollo}, {F{\"o}rster Schreiber}, {Renzini}, {Dekel}, {Genzel}, {Lang}, {Lilly}, {Mancini}, {Onodera}, {Tacconi}, {Wuyts}, \& {Zamorani}}]{Tacchella_2018}
{Tacchella}, S., {Carollo}, C.~M., {F{\"o}rster Schreiber}, N.~M., {et~al.} 2018, \apj, 859, 56

\bibitem[{{Tacchella} {et~al.}(2015{\natexlab{a}}){Tacchella}, {Carollo}, {Renzini}, {F{\"o}rster Schreiber}, {Lang}, {Wuyts}, {Cresci}, {Dekel}, {Genzel}, {Lilly}, {Mancini}, {Newman}, {Onodera}, {Shapley}, {Tacconi}, {Woo}, \& {Zamorani}}]{Tacchella_2015b}
{Tacchella}, S., {Carollo}, C.~M., {Renzini}, A., {et~al.} 2015{\natexlab{a}}, Science, 348, 314

\bibitem[{{Tacchella} {et~al.}(2016){Tacchella}, {Dekel}, {Carollo}, {Ceverino}, {DeGraf}, {Lapiner}, {Mandelker}, \& {Primack}}]{Tacchella_2016b}
{Tacchella}, S., {Dekel}, A., {Carollo}, C.~M., {et~al.} 2016, \mnras, 458, 242

\bibitem[{{Tacchella} {et~al.}(2015{\natexlab{b}}){Tacchella}, {Lang}, {Carollo}, {F{\"o}rster Schreiber}, {Renzini}, {Shapley}, {Wuyts}, {Cresci}, {Genzel}, {Lilly}, {Mancini}, {Newman}, {Tacconi}, {Zamorani}, {Davies}, {Kurk}, \& {Pozzetti}}]{Tacchella_2015a}
{Tacchella}, S., {Lang}, P., {Carollo}, C.~M., {et~al.} 2015{\natexlab{b}}, \apj, 802, 101

\bibitem[{{Thomas} {et~al.}(1999){Thomas}, {Greggio}, \& {Bender}}]{Thomas_1999}
{Thomas}, D., {Greggio}, L., \& {Bender}, R. 1999, \mnras, 302, 537

\bibitem[{{Tinsley}(1979)}]{Tinsley_1979}
{Tinsley}, B.~M. 1979, \apj, 229, 1046

\bibitem[{{Tripodi} {et~al.}(2024){Tripodi}, {D'Eugenio}, {Maiolino}, {Curti}, {Scholtz}, {Tacchella}, {Marconcini}, {Bunker}, {Trussler}, {Cameron}, {Arribas}, {Baker}, {Brada{\v{c}}}, {Carniani}, {Charlot}, {Ji}, {Ji}, {Robertson}, {{\"U}bler}, {Venturi}, {Willmer}, \& {Witstok}}]{Tripodi_2024}
{Tripodi}, R., {D'Eugenio}, F., {Maiolino}, R., {et~al.} 2024, \aap, 692, A184

\bibitem[{{Trujillo} {et~al.}(2009){Trujillo}, {Cenarro}, {de Lorenzo-C{\'a}ceres}, {Vazdekis}, {de la Rosa}, \& {Cava}}]{Trujillo_2009}
{Trujillo}, I., {Cenarro}, A.~J., {de Lorenzo-C{\'a}ceres}, A., {et~al.} 2009, \apjl, 692, L118

\bibitem[{{Trujillo} {et~al.}(2004){Trujillo}, {Rudnick}, {Rix}, {Labb{\'e}}, {Franx}, {Daddi}, {van Dokkum}, {F{\"o}rster Schreiber}, {Kuijken}, {Moorwood}, {R{\"o}ttgering}, {van der Wel}, {van der Werf}, \& {van Starkenburg}}]{Trujillo_2004}
{Trujillo}, I., {Rudnick}, G., {Rix}, H.-W., {et~al.} 2004, \apj, 604, 521

\bibitem[{{Trussler} {et~al.}(2020){Trussler}, {Maiolino}, {Maraston}, {Peng}, {Thomas}, {Goddard}, \& {Lian}}]{Trussler_2020}
{Trussler}, J., {Maiolino}, R., {Maraston}, C., {et~al.} 2020, \mnras, 491, 5406

\bibitem[{{Valentino} {et~al.}(2023){Valentino}, {Brammer}, {Gould}, {Kokorev}, {Fujimoto}, {Jespersen}, {Vijayan}, {Weaver}, {Ito}, {Tanaka}, {Ilbert}, {Magdis}, {Whitaker}, {Faisst}, {Gallazzi}, {Gillman}, {Gim{\'e}nez-Arteaga}, {G{\'o}mez-Guijarro}, {Kubo}, {Heintz}, {Hirschmann}, {Oesch}, {Onodera}, {Rizzo}, {Lee}, {Strait}, \& {Toft}}]{Valentino_2023}
{Valentino}, F., {Brammer}, G., {Gould}, K. M.~L., {et~al.} 2023, \apj, 947, 20

\bibitem[{{van de Sande} {et~al.}(2013){van de Sande}, {Kriek}, {Franx}, {van Dokkum}, {Bezanson}, {Bouwens}, {Quadri}, {Rix}, \& {Skelton}}]{van_de_Sande_2013}
{van de Sande}, J., {Kriek}, M., {Franx}, M., {et~al.} 2013, \apj, 771, 85

\bibitem[{{van der Wel} {et~al.}(2014){van der Wel}, {Franx}, {van Dokkum}, {Skelton}, {Momcheva}, {Whitaker}, {Brammer}, {Bell}, {Rix}, {Wuyts}, {Ferguson}, {Holden}, {Barro}, {Koekemoer}, {Chang}, {McGrath}, {H{\"a}ussler}, {Dekel}, {Behroozi}, {Fumagalli}, {Leja}, {Lundgren}, {Maseda}, {Nelson}, {Wake}, {Patel}, {Labb{\'e}}, {Faber}, {Grogin}, \& {Kocevski}}]{van_der_Wel_2014}
{van der Wel}, A., {Franx}, M., {van Dokkum}, P.~G., {et~al.} 2014, \apj, 788, 28

\bibitem[{{van der Wel} {et~al.}(2008){van der Wel}, {Holden}, {Zirm}, {Franx}, {Rettura}, {Illingworth}, \& {Ford}}]{van_der_wel_2008}
{van der Wel}, A., {Holden}, B.~P., {Zirm}, A.~W., {et~al.} 2008, \apj, 688, 48

\bibitem[{{van der Wel} {et~al.}(2024){van der Wel}, {Martorano}, {H{\"a}u{\ss}ler}, {Nedkova}, {Miller}, {Brammer}, {van de Ven}, {Leja}, {Bezanson}, {Muzzin}, {Marchesini}, {de Graaff}, {Nelson}, {Kriek}, {Bell}, \& {Franx}}]{van_der_wel_2024}
{van der Wel}, A., {Martorano}, M., {H{\"a}u{\ss}ler}, B., {et~al.} 2024, \apj, 960, 53

\bibitem[{{van Dokkum} \& {Conroy}(2024)}]{van_dokkum_2024}
{van Dokkum}, P. \& {Conroy}, C. 2024, \apjl, 973, L32

\bibitem[{{van Dokkum} {et~al.}(2017){van Dokkum}, {Conroy}, {Villaume}, {Brodie}, \& {Romanowsky}}]{van_dokkum_2017}
{van Dokkum}, P., {Conroy}, C., {Villaume}, A., {Brodie}, J., \& {Romanowsky}, A.~J. 2017, \apj, 841, 68

\bibitem[{{van Dokkum} {et~al.}(2014){van Dokkum}, {Bezanson}, {van der Wel}, {Nelson}, {Momcheva}, {Skelton}, {Whitaker}, {Brammer}, {Conroy}, {F{\"o}rster Schreiber}, {Fumagalli}, {Kriek}, {Labb{\'e}}, {Leja}, {Marchesini}, {Muzzin}, {Oesch}, \& {Wuyts}}]{van_dokkum_2014}
{van Dokkum}, P.~G., {Bezanson}, R., {van der Wel}, A., {et~al.} 2014, \apj, 791, 45

\bibitem[{{van Dokkum} \& {Franx}(2001)}]{van_dokkum_2001}
{van Dokkum}, P.~G. \& {Franx}, M. 2001, \apj, 553, 90

\bibitem[{{van Dokkum} {et~al.}(2010){van Dokkum}, {Whitaker}, {Brammer}, {Franx}, {Kriek}, {Labb{\'e}}, {Marchesini}, {Quadri}, {Bezanson}, {Illingworth}, {Muzzin}, {Rudnick}, {Tal}, \& {Wake}}]{van_dokkum_2010}
{van Dokkum}, P.~G., {Whitaker}, K.~E., {Brammer}, G., {et~al.} 2010, \apj, 709, 1018

\bibitem[{{Villaume} {et~al.}(2017){Villaume}, {Conroy}, {Johnson}, {Rayner}, {Mann}, \& {van Dokkum}}]{Villaume_2017}
{Villaume}, A., {Conroy}, C., {Johnson}, B., {et~al.} 2017, \apjs, 230, 23

\bibitem[{Virtanen {et~al.}(2020)Virtanen, Gommers, Oliphant, Haberland, Reddy, Cournapeau, Burovski, Peterson, Weckesser, Bright, {van der Walt}, Brett, Wilson, Millman, Mayorov, Nelson, Jones, Kern, Larson, Carey, Polat, Feng, Moore, {VanderPlas}, Laxalde, Perktold, Cimrman, Henriksen, Quintero, Harris, Archibald, Ribeiro, Pedregosa, {van Mulbregt}, \& {SciPy 1.0 Contributors}}]{2020SciPy-NMeth}
Virtanen, P., Gommers, R., Oliphant, T.~E., {et~al.} 2020, Nature Methods, 17, 261

\bibitem[{{Wellons} {et~al.}(2015){Wellons}, {Torrey}, {Ma}, {Rodriguez-Gomez}, {Vogelsberger}, {Kriek}, {van Dokkum}, {Nelson}, {Genel}, {Pillepich}, {Springel}, {Sijacki}, {Snyder}, {Nelson}, {Sales}, \& {Hernquist}}]{Wellons_2015}
{Wellons}, S., {Torrey}, P., {Ma}, C.-P., {et~al.} 2015, \mnras, 449, 361

\bibitem[{{Whitaker} {et~al.}(2012){Whitaker}, {Kriek}, {van Dokkum}, {Bezanson}, {Brammer}, {Franx}, \& {Labb{\'e}}}]{Whitaker_2012}
{Whitaker}, K.~E., {Kriek}, M., {van Dokkum}, P.~G., {et~al.} 2012, \apj, 745, 179

\bibitem[{{Worthey}(1994)}]{Worthey_1994}
{Worthey}, G. 1994, \apjs, 95, 107

\bibitem[{{Wuyts} {et~al.}(2010){Wuyts}, {Cox}, {Hayward}, {Franx}, {Hernquist}, {Hopkins}, {Jonsson}, \& {van Dokkum}}]{Wuyts_2010}
{Wuyts}, S., {Cox}, T.~J., {Hayward}, C.~C., {et~al.} 2010, \apj, 722, 1666

\bibitem[{{Zhuang} {et~al.}(2023){Zhuang}, {Leethochawalit}, {Kirby}, {Nightingale}, {Steidel}, {Glazebrook}, {Barone}, {Skobe}, {Sweet}, {Nanayakkara}, {Allen}, {Vasan}, {Jones}, {Kacprzak}, {Tran}, \& {Jacobs}}]{Zhuang_2023}
{Zhuang}, Z., {Leethochawalit}, N., {Kirby}, E.~N., {et~al.} 2023, \apj, 948, 132

\bibitem[{{Zibetti} {et~al.}(2020){Zibetti}, {Gallazzi}, {Hirschmann}, {Consolandi}, {Falc{\'o}n-Barroso}, {van de Ven}, \& {Lyubenova}}]{Zibetti_2020}
{Zibetti}, S., {Gallazzi}, A.~R., {Hirschmann}, M., {et~al.} 2020, \mnras, 491, 3562

\bibitem[{{Zolotov} {et~al.}(2015){Zolotov}, {Dekel}, {Mandelker}, {Tweed}, {Inoue}, {DeGraf}, {Ceverino}, {Primack}, {Barro}, \& {Faber}}]{Zolotov_2015}
{Zolotov}, A., {Dekel}, A., {Mandelker}, N., {et~al.} 2015, \mnras, 450, 2327

\bibitem[{{Zolotov} {et~al.}(2010){Zolotov}, {Willman}, {Brooks}, {Governato}, {Hogg}, {Shen}, \& {Wadsley}}]{Zolotov_2010}
{Zolotov}, A., {Willman}, B., {Brooks}, A.~M., {et~al.} 2010, \apj, 721, 738

\end{thebibliography}

\begin{appendix}
\section{Mock observations}\label{sec:mocks}
To determine our S/N limits (see Section~\ref{sec:data_sample}) and fitting strategy (see Section~\ref{sec:alf}) we performed test fits with simulated \textit{JWST}-SUSPENSE observations, as discussed in Section~\ref{sec:data_sample}.  These tests are very similar to the ones performed in \cite{Cheng_2024}.  We generated mock spectra using the model grids in \textsc{alf$\alpha$}.  We used the integrated fit results from \cite{Beverage_suspense} to set the values of the stellar population parameters (age, velocity dispersion, and elemental abundances) to the values of an example galaxy (127154) from our sample.  We then scaled these mock spectra to have different S/N, using the noise spectrum from the real observation of 127154.  We scaled the noise to the desired S/N between $4520-4820$ \AA\ (the common wavelength region covered by our entire sample, see Section~\ref{sec:data_sample}) and added the noise that we randomly sampled from a Gaussian distribution to the mock flux.

\begin{figure*}
    \centering
    \includegraphics[width=0.8\textwidth]{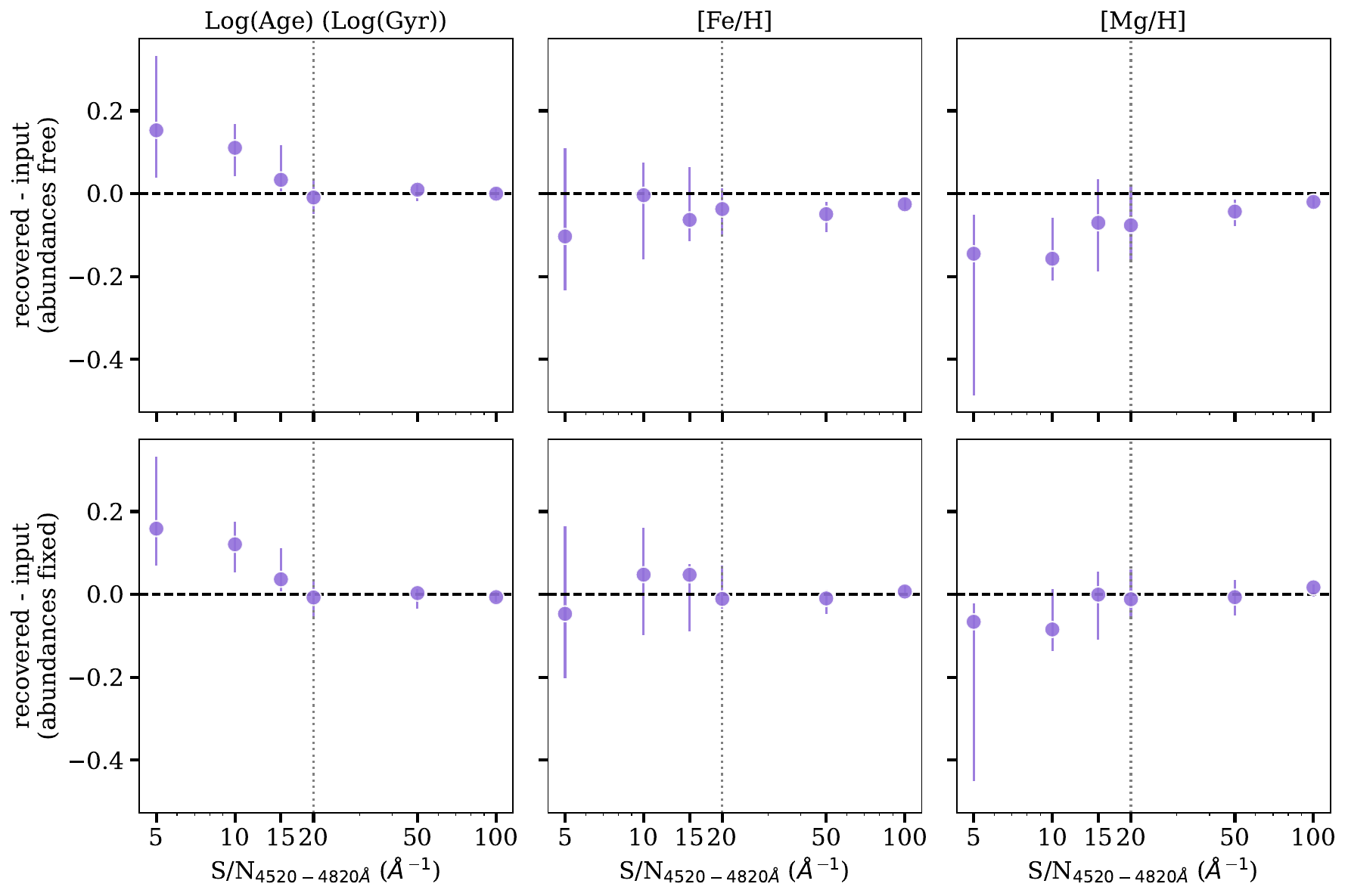}
    \caption{Tests with mock \textit{JWST}-SUSPENSE observations.  In each panel, we show the difference between the recovered and input parameters.  We show true recovery (a difference of zero) with the black, dashed line and we show the S/N limit that we use in the body of the paper as the vertical, dotted, grey line.  We show $\log(\rm Age)$ in the left column, [Fe/H] in the middle column, and [Mg/H] in the right column.  In the top row, we left all stellar population parameters free in the \textsc{alf$\alpha$} fits.  In the bottom row, we fixed all elemental abundances other than [Fe/H] and [Mg/H].  All three parameters (age, Fe, and Mg) can be reliably recovered at a S/N $\gtrsim20$ \AA\ between $4520-4820$ \AA.  Moreover, the recovery is much more accurate when the remaining elemental abundances are fixed.  Thus, the fitting strategy shown in the bottom row is the one that we use throughout the paper.}
    \label{fig:mocks}
\end{figure*}

The purpose of these tests was twofold: (i) we tested the S/N that we needed to recover ages, Fe-, and Mg-abundances, and (ii) we tested whether we needed to fix the remaining elemental abundances to their integrated values, or whether we could leave them free.  We show the results of these tests in Figure~\ref{fig:mocks}.  We generated 10 mocks in each of six S/N bins and fit each individual mock with \textsc{alf$\alpha$}.  In each panel of Figure~\ref{fig:mocks}, we took the difference between the recovered and input parameters from each fit and show the medians and $1\sigma$ errors.  In the top row, we show a test where we leave all stellar population parameters free in the fits.  In the bottom row, we show a test where we fix all elemental abundances other than Fe and Mg to their integrated values from \cite{Beverage_suspense}.  This test demonstrates that we require a S/N $\gtrsim 20$ \AA$^{-1}$ between $4520 - 4820$ \AA\ in order to reliably recover age, Fe, and Mg.  Additionally, the recovery of these three parameters is much more accurate when we fix the elemental abundances to their integrated values.  In the top panel, it is difficult to recover the input parameters when all parameters are left free, even with very high S/N data. 

\section{Wiggle correction algorithm}\label{sec:appendix_wiggles}
\begin{figure*}
    \centering
    \includegraphics[width=\textwidth]{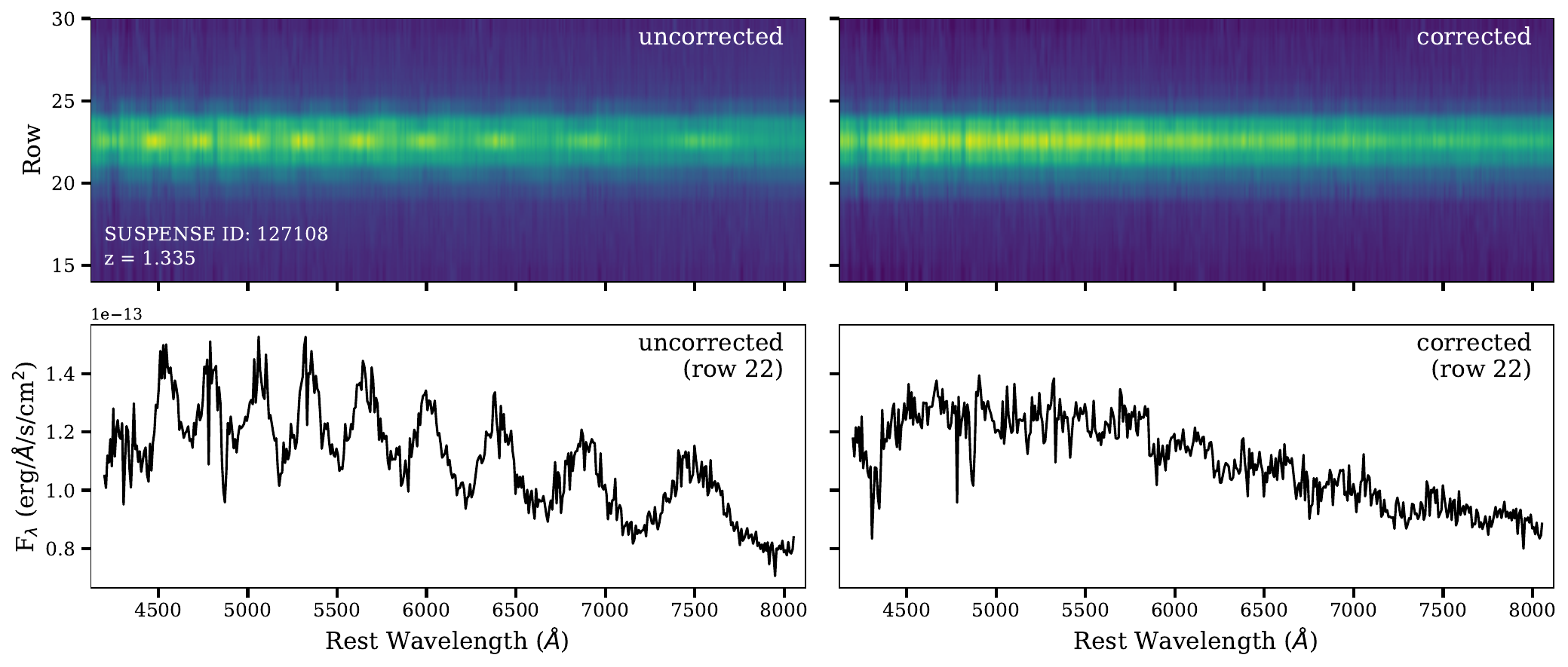}
    \caption{Example of the effect of our resampling noise correction for one object in our sample (object ID 127108).  \textit{Top}: Two-dimensional spectra.  The uncorrected spectrum is shown on the left, where bright variations in the flux can be seen across the wavelength range.  The corrected spectrum is shown on the right, where the flux is much more uniform across the wavelength range.  \textit{Bottom}: Extracted 1D rows, median-binned by 3 pixels.  In the left panel, we extracted the brightest row from the uncorrected 2D spectrum.  The bright variations in the flux seen in the top panel manifest as a sinusoidal flux variation in the 1D row.  In the right panel, we extracted the brightest row from the corrected 2D spectrum, where this sinusoidal flux variation has been removed.}
    \label{fig:wiggle_before_after}
\end{figure*}
As discussed in Section~\ref{sec:wiggles}, spatially resolved \textit{JWST} observations are affected by undersampling of the PSF.  Algorithms to correct these wiggles in NIRSpec IFU data have been introduced in \cite{Perna_2023} and in the \textsc{WiCKED} Python package \citep{wicked}.  However, at the time of writing, this study is the first one that tries to correct the wiggles in spatially resolved NIRSpec-MSA spectra.  Thus, we implemented an algorithm that is heavily based on the ones presented in \cite{Perna_2023} and \textsc{WiCKED}, but we combined and customised these routines for our specific case.  We summarise the steps in our algorithm here.

We first fit the wiggles using the spectrum extracted from the row with the highest S/N.  To obtain the shape of the wiggle, we masked strong emission lines.  Then, taking the integrated spectrum as a reference spectrum, we divided the brightest row by this reference, with both the reference and individual row spectra normalised by the value between $4480-4520$ \AA\ (see \citealt{Slob_2024}).  We broke the wiggle spectrum up into small regions (non-uniform in length) that covered the entire wavelength range and used a sinusoidal function to model the wiggle in each chunk.  We iterated this process to find both the best chunk sizes and best-fitting sinusoidal models by minimising the $\chi^2$ value over the entire wiggle spectrum.  We divided this wiggle model from the spectral row to produce a corrected spectrum in the brightest row.  For objects with a detector gap, we fit the wiggle on each side of the gap independently (if the spectrum was long enough for a wiggle to be detected visually). 

For the subsequent spectral rows, we implemented the Fourier transform algorithm from \textsc{WiCKED} to identify which rows contained a wiggle.  In particular, for rows near the edges, the spectrum is dominated by noise and the S/N is too low to detect a wiggle.  Additionally, there are certain brighter rows that are not as strongly affected by the undersampling and spectral rectification, and thus do not require wiggle correction.  For the rows that contained wiggles, we used the chunks and wiggle model fit in the brightest row as a prior to constrain the wiggle in the remaining rows.  We visually checked the fits in each row.  In cases where these priors did not produce a good fit, we re-fit the rows without a prior.  Similar to the brightest row, we divided the wiggle model for each row out of each respective spectral row to obtain a corrected spectrum.  We ensured that flux was conserved during the wiggle correction procedure.  We show an example of the wiggles and our correction in Fig.~\ref{fig:wiggle_before_after}.  

 \begin{figure*}
    \centering
    \includegraphics[width=0.8\textwidth]{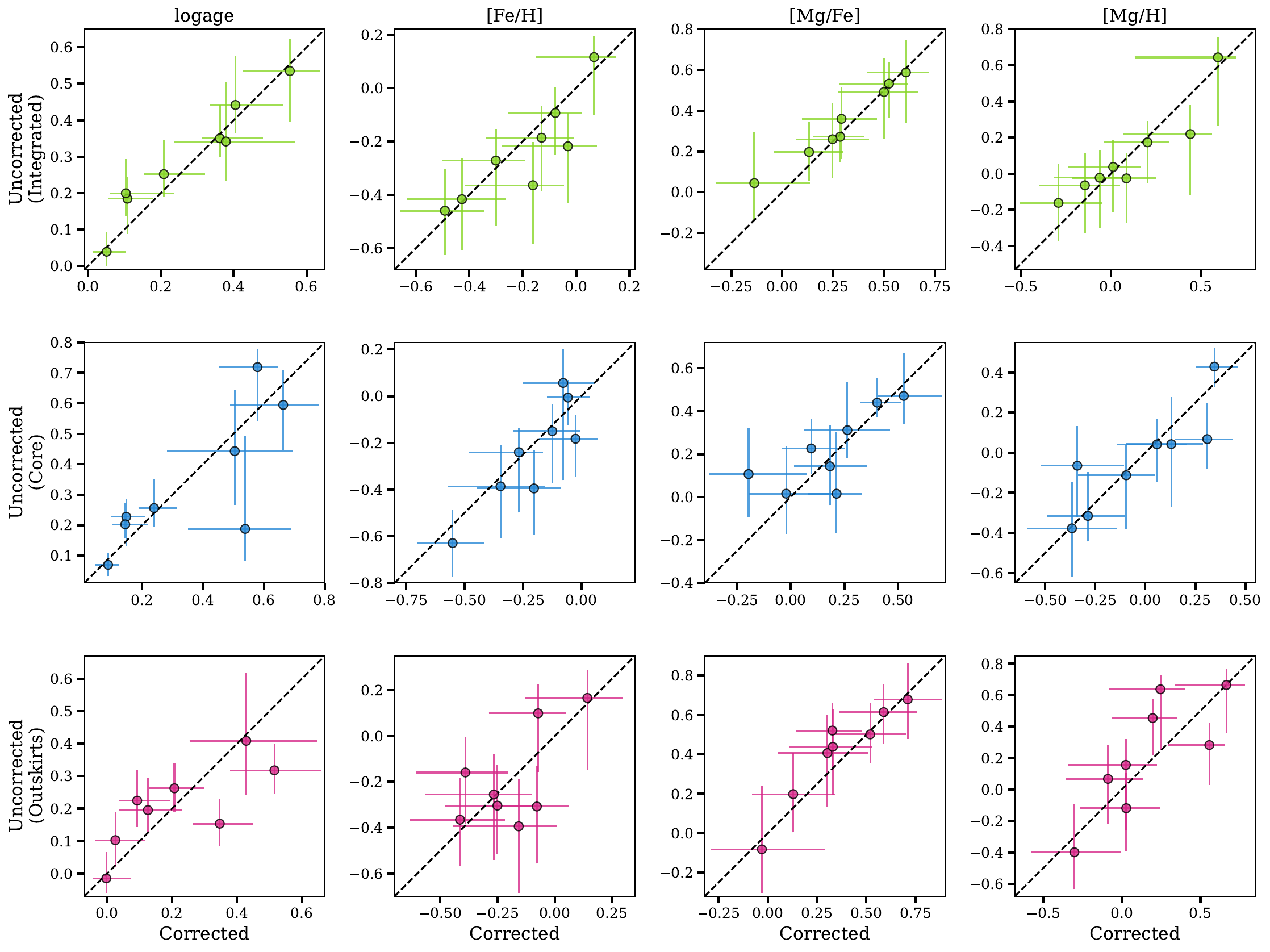}
    \caption{One-to-one plots comparing our fiducial (wiggle-corrected) fits to fits where we do not correct the wiggles.  The integrated fits are shown in the top row, the core fits are shown in the middle row, and the outskirt fits are shown in the bottom row.  We compare age in the left column, [Fe/H] in the second, [Mg/Fe] in the third column, and [Mg/H] in the right column.  Uncorrected fit results are shown on the $y$-axes and fiducial fit results are shown on the $x$-axes.}
    \label{fig:wiggle_param_comparison}
\end{figure*}

We tested the impact of the wiggles and the wiggle correction on our results by extracting and fitting the uncorrected spectra in the same way as the corrected spectra for the galaxies in our sample.  In Figure~\ref{fig:wiggle_param_comparison}, we show one-to-one plots comparing our fitted ages (left column), [Fe/H] values (middle column), and [Mg/Fe] values (right column) for our integrated (top row), core (middle row), and outskirt (bottom row) spectra.  The integrated fits are largely consistent, showing that the wiggles do not have a significant impact on integrated spectra, although we note that there is still some scatter.  The core and outskirt fits are also relatively consistent but have more scatter around the one-to-one relation compared to the integrated fits.  We also note that \cite{wicked} show that this wiggle correction algorithm does not affect the true equivalent widths of absorption features, and only acts to correct the broad, sinusoidal artefact across broad wavelength regions.    

\end{appendix}
\end{document}